\documentclass[usenatbib]{mn2e}

\usepackage{epsfig}
\usepackage{psfig}

\newcommand{\dpem}{$D^\prime(P_i,e_i,M_i)$}
\newcommand{\mjup}{M$_{\rm Jup}$}
\begin{document}

\title[Selection Functions in Planet Searches]{Selection Functions in Doppler Planet Searches}
\author[S.~J.~O'Toole et al.]{S.~J.~O'Toole,$^{1,3}$
C.~G.~Tinney,$^2$
H.~R.~A.~Jones,$^3$
R.~P.~Butler,$^4$,
G.~W.~Marcy,$^{5,6}$ \newauthor
B.~Carter,$^7$ and
J.~Bailey$^8$ \\
$^1$Anglo-Australian Observatory, PO Box 296, Epping 1710, Australia \\
$^2$Department of Astrophysics, School of Physics, University of NSW,
  2052, Australia \\
$^3$Centre for Astrophysics Research, University of Hertfordshire,
Hatfield, AL 10 9AB, UK \\
$^4$Department of Terrestrial Magnetism, Carnegie Institution of
Washington, 5241 Broad Branch Road NW, Washington DC, USA 20015-1305
\\
$^5$Department of Astronomy, University of California, Berkeley, CA
USA 94720 \\
$^6$Department of Physics and Astronomy, San Francisco State
University, San Francisco, CA, USA 94132 \\
$^7$Faculty of Sciences, University of Southern Queensland, Toowoomba,
Queensland 4350, Australia \\
$^8$Physics Department, Macquarie University, Sydney,
NSW 2109, Australia}

\maketitle
\begin{abstract}

We present a preliminary analysis of the sensitivity of
Anglo-Australian Planet Search data to the orbital parameters of
extrasolar planets. To do so, we have developed new tools for the
automatic analysis of large-scale simulations of Doppler velocity
planet search data. One of these tools is the 2-Dimensional Keplerian
Lomb-Scargle periodogram, that enables the straightforward detection
of exoplanets with high eccentricities (something the standard
Lomb-Scargle periodogram routinely fails to do). We used this
technique to re-determine the orbital parameters of HD\,20782b, with
one of the highest known exoplanet eccentricities
($e=0.97\pm0.01$). We also derive a set of detection criteria that do
not depend on the distribution functions of fitted Keplerian orbital
parameters (which we show are non-Gaussian with pronounced, extended
wings). Using these tools, we examine the selection functions in
orbital period, eccentricity and planet mass of Anglo-Australian
Planet Search data for three planets with large-scale Monte Carlo-like
simulations. We find that the detectability of exoplanets declines at
high eccentricities. However, we also find that exoplanet
detectability is a strong function of epoch-to-epoch data quality,
number of observations, and period sampling. This strongly suggests
that simple parametrisations of the detectability of exoplanets based
on ``whole-of-survey'' metrics may not be accurate. We have derived
empirical relationships between the uncertainty estimates for orbital
parameters that are derived from least-squares Keplerian fits to our
simulations, and the {\em true} 99\% limits for the errors in those
parameters, which are larger than equivalent Gaussian limits by
factors of 5-10. We quantify the rate at which false positives are
made by our detection criteria, and find that they do not
significantly affect our final conclusions. And finally, we find that
there is a bias against measuring near-zero eccentricities, which
becomes more significant in small, or low signal-to-noise-ratio, data
sets.

\end{abstract}

\section{Introduction}
\label{sec:intro}

Extra-solar planet detection using the Doppler method has played a
dominant role in placing this field at the heart of astronomical
research. The advances made in this field have been primarily due to
the significantly
increased stability of high-resolution spectrographs, and improved
techniques for calibrating residual spectrograph variations. An
excellent example of this can be seen in the Anglo-Australian Planet
Search (AAPS), where a long-term velocity precision of 3\,m\,s$^{-1}$
over the initial $\sim$\,8 years of observation \citep{TBM05}, has
been improved in recent years to better than 2\,m\,s$^{-1}$
\citep[e.g.][]{OBT07}.

With the time baselines of Doppler surveys now approaching (or
exceeding) a decade, and routine velocity precisions approaching
1\,m\,s$^{-1}$, we are now in a position to ask, and answer, critical
questions about the underlying distributions of exoplanet
parameters. Questions such as, how common are gas-giant planets in
Jupiter-like orbits (i.e.\ $\sim$5\,AU near circular orbits)? How
common are gas-giant planets in Earth-like orbits, likely to host
habitable terrestrial satellites (i.e.\ $\sim$1\,AU near circular
orbits)? And, how common are low-mass planets in close orbits
(i.e. $\sim <$\,0.3\,AU with M\,$\sin i<10\,M_{\mathrm Earth}$)?

To answer these questions, we must first characterise and quantify the
selection effects that are present in our observations. It is clear
that the Doppler velocity planet searches {\em have} inherent
observational biases. For example, \citet{GSM-O05} have noted that
there is a sharp cut-off at periods of $\sim$\,3 days for planets
detected in RV surveys, while transit surveys have found the majority
of their planets at periods below this. We need to know, in detail,
what orbital parameter space Doppler surveys probe, and how their
sensitivity varies over that orbital parameter space.

Analysis of Doppler survey selection effects have, to date, been
forced to make a variety of simplifying assumptions.  \citet{NCL05},
for example, examined the detectability of short-period, close-in
planets, and so were able to ignore the complicating effects of
eccentricity. They derived an empirical relation for detectability as
a function of the number of observations and data
quality. \citet{WEC06} investigated the detectability of exoplanets in
their data using simulations at $e=0$ and 0.6, enabling them to
determine that there \emph{are} selection effects at high
eccentricity, but not to quantify them across the full range of
eccentricities which exoplanets have been found to display.  The effects
of eccentricity were considered by \citet{Cumming04}, who derived
empirical relationships for velocity thresholds relying on an
$F$-statistic for two different cases: when the orbital period is
shorter than the time-span of the observations, and when it is
longer. More recently, (in an expansion of work begun in
\citet{CMB99}), \citet{CBM08} used this technique to derive detection
thresholds, determine selection effects and measure the incompleteness
of Keck Planet Search data, in order to investigate the exoplanetary
minimum mass and orbital period distributions present in that
data. However, the analytical method used in these studies makes a
number of simplifying assumptions: that individual velocity uncertainties can
be represented with a Gaussian distribution; that observations are
evenly spaced; and that the number of independent periods probed by a
data set can be quantified in a meaningful way
\citep{MBV05}. Unfortunately, real observations violate all three of
these assumptions.

In this paper, therefore, we lay the groundwork for an investigation
of the full range of physically interesting exoplanet parameters that
Doppler data can probe (period $P$,  eccentricity $e$ and minimum
planet mass $M\sin i$) using star-by-star, epoch-by-epoch Monte Carlo
simulations, in an effort to understand what our Doppler data are
telling us about the orbital parameters of exoplanets, while making as
few assumptions as possible.  We introduce a set of automated planet
detection criteria and combine it with large-scale simulations of
Keplerian orbits for each star observed, to determine the sensitivity
of our ``as observed'' data.

\subsection{Observations, sampling and data quality}
\label{sub:obs}

Objects in the AAPS catalogue are listed in \citet{JBM02} and
\citet{TBM03}. Details of our observing program are described
in more detail elsewhere \citep{BTM01}. Briefly, the
data are taken using the University College London Echelle
Spectrograph mounted at the coud\'e focus of the
Anglo-Australian Telescope (AAT). An iodine absorption cell is placed
in the beam, imprinting a forest of molecular iodine absorption lines
onto the stellar spectrum. These lines are used as a wavelength
reference to derive high-precision velocities as described in
\citet{BMW96}.

The target stars of the AAPS (in common with most Doppler search
targets) are observed in a non-uniform way. First, observing runs are
scheduled in blocks spread unevenly throughout a semester, which
necessarily needs to non-ideal (ie. non-logarithmic) period sampling.
Second, the weather during each block of observations affects the
time-sampling of data as well, with velocity precisions generally
being poorer in poor weather conditions, and with bright objects
tending to be generally observed more often, when conditions are
poor. (Note that in this paper we use the median measurement
uncertainty of a given set of observations as an indicator of the data
quality.)

Finally, large amounts of data tend to be acquired for objects
where a planet is thought to exist, and smaller amounts of
data are acquired for stars where planets are thought unlikely
(or where a possible planet has period longer than the current data string).
As a result ``high priority'' targets get observed more densely.
As a result of all these effects, time-sampling can deviate markedly
from the {\em ideal} of uniform logarithmic sampling in period space.

After each observing block is completed, the data are processed to
update a database of velocities. This is analysed periodically to look
for objects showing significant variability or periodicity. These get
promoted to the ``higher priority'' status described above. This
prioritisation analysis has been done to date using a Lomb-Scargle
periodogram, Keplerian fitting based on the most significant periods,
the determination of False Alarm Probabilities \citep{MBV05}, and the
application of simple tests asking ``Have we seen at least one
period?'' and ``Do subsequent data obtained from a high priority
object match the prediction from initial fits?'' This prioritisation
maximises the rate at which exoplanets can be extracted from our data,
but means that our survey database (like those of most Doppler planet
searches) is quite non-uniform. Simulation of the as-observed data
sets for all stars in our database is therefore the only way to
quantify the non-uniform selection effects inherent in Doppler
velocity planet searches.

\section{The 2D Keplerian Lomb-Scargle Periodogram}
\label{sec:2dkls}

A tool commonly used for detecting variability in light and radial
velocity curves is the Lomb-Scargle (LS) periodogram
\citep{Lomb76,Scargle82}. It involves determining, as a function of
frequency, the difference between the $\chi^2$ of a sinusoid fit to
data and the $\chi^2$ of a constant fit (with the the resulting
``power'' being normalised in some way). There are well developed
statistics surrounding LS power, allowing significance values to be
attributed to possible detections \citep[see e.g.][for a
discussion]{Cumming04}. In most cases where the LS periodogram is
applied (e.g. in most areas of stellar pulsation and close binarity)
the signal under study is approximately sinusoidal, and so the LS
periodogram is applied appropriately.

The LS periodogram is now increasingly being used in Doppler velocity
planet searches where, however, circular orbits (giving rise to
sinusoidal velocity curves) are not common \citep{BWM06}. It
therefore makes more sense to fit Keplerians to data instead of
sinusoids as discussed by \citet{Cumming04}. As orbital eccentricity
is an important parameter in a Keplerian function, we have expanded the
traditional LS periodogram to include two dimensions -- that is, to examine
power as a function of both period and eccentricity.
We call this the 2D Keplerian LS (2DKLS) periodogram. The method we
use to calculate the 2DKLS periodogram was briefly discussed in
\citet{OBT07} and is described in more detail below.

\subsection{Method}
\label{sub:method}

The 2DKLS is an extension of the traditional
Lomb-Scargle periodogram, where we vary period {\em and} eccentricity
in the calculation of power. (While the argument of pericenter,
$\omega$, is also important in determining the \emph{shape} of the
velocity curve, it does \emph{not} impact the orbital period
measurement in the same way as eccentricity.)
This is also an extension of the one dimensional Keplerian
Lomb-Scargle periodogram introduced by \citet{Cumming04}. We
find, however, that not fixing eccentricity, while more efficient
computationally, allows for possible non-physical values (i.e. outside
the range 0-1). The ``smoothness'' of the periodogram also depends a
lot more on the initial guesses for the free parameters.
We use a grid of fixed periods and eccentricities to calculate the
2DKLS, with $e=0-0.98$ in steps of 0.01, and periods
on a logarithmic scale from 1 day up to the maximum possible period of
interest for that data sequence (in most cases 4500 days for current
AAPS data), on a spacing of $10^{-3}$ in $\log_{\mathrm{10}}P$. A
Keplerian described by Equation \ref{eq:kepler} is then fitted to the
data using a non-linear least squares fitting routine with
Levenberg-Marquardt minimisation from \citet{Press86}. 

\begin{equation}
V_r(t)=K[\cos(\omega+\nu(t))+e\cos\omega]+V_{\mathrm{0}}
\label{eq:kepler}
\end{equation}

Here $K$ is the semi-amplitude, $\nu(t)$ is the true anomaly involving
implicit dependence on the orbital period $P$ and the time of
periastron passage $T_{\mathrm{p}}$, and $V_{\mathrm{0}}$ is the
velocity zero-point. The true anomaly is derived by solving Kepler's
equation $M(t)=E(t)-e\sin E(t)$ where $E(t)$ is the eccentric anomaly and
$M(t)=2\pi t/P$ is the mean anomaly. The power, $z(P,e)$, is determined using
$z(P,e)=\Delta\chi^2/4=(\chi^2_{\mathrm{mean}}-\chi^2_{\mathrm{Kep}})/4$,
where $\chi^2_{\mathrm{Kep}}$ is the goodness-of-fit for the best fit
Keplerian model, and $\chi^2_{\mathrm{mean}}$ is the equivalent for a
constant fit to the data. For each value of $P$ and $e$ we find the
values of the remaining parameters that minimise
$\chi^2_{\mathrm{Kep}}$ and therefore maximise $z(P,e)$. As discussed
by \citet{Cumming04}, when the noise level is not known in advance
(i.e. for observations) $z(P,e)$ must be normalised, in the 2DKLS case
by $\chi^2_{\mathrm{Kep}}$. This form of the 2DKLS was implemented by
\citet{OBT07} and is used in the next section. For the purposes of the
simulations in this paper (described in Section \ref{sec:simul}),
the power is not normalised, because the noise level is an input
parameter and is therefore known in advance.

\begin{figure}
  \begin{center}
    \leavevmode
    \leavevmode
    \psfig{file=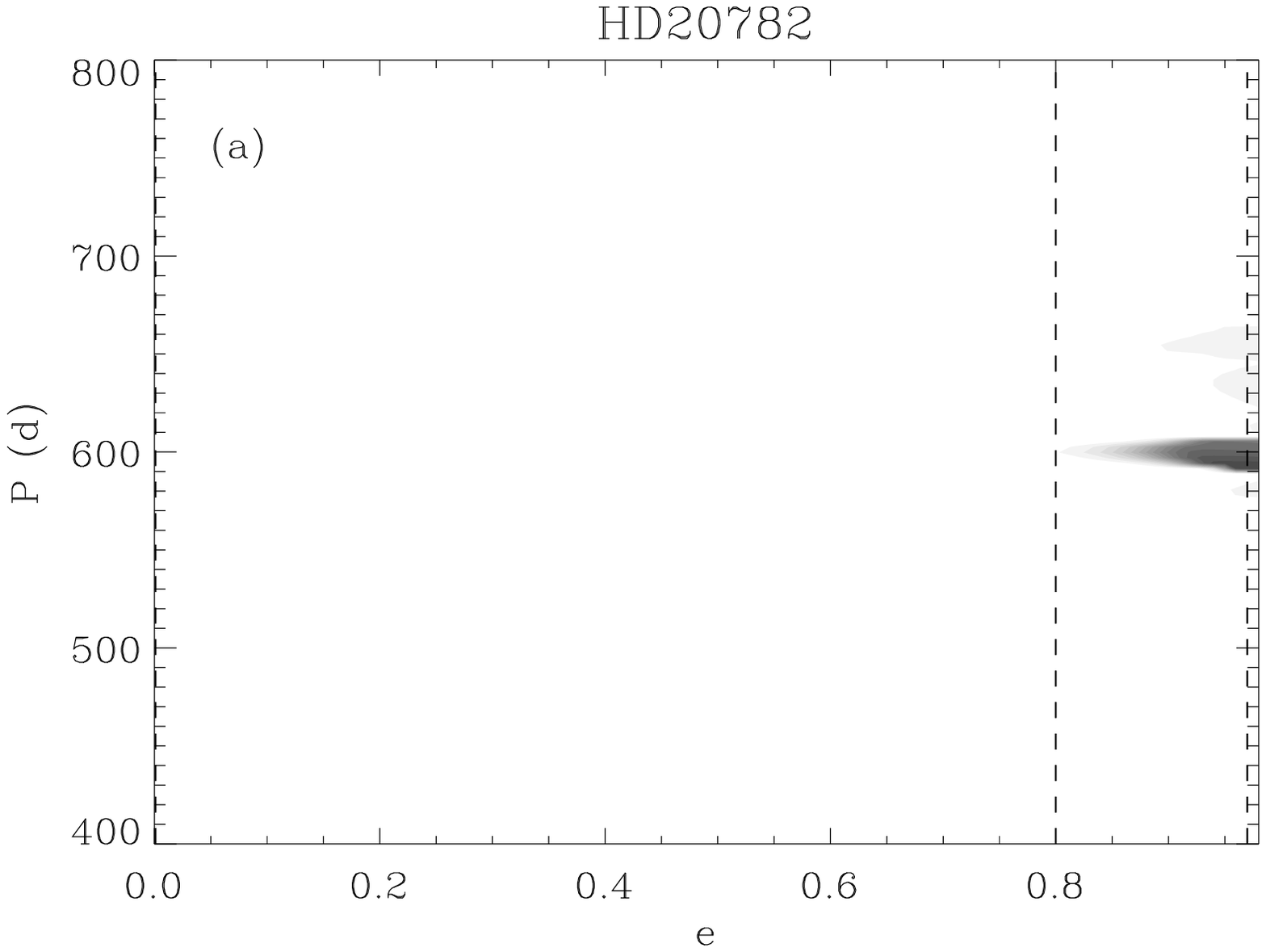,scale=0.47}
    \epsfig{file=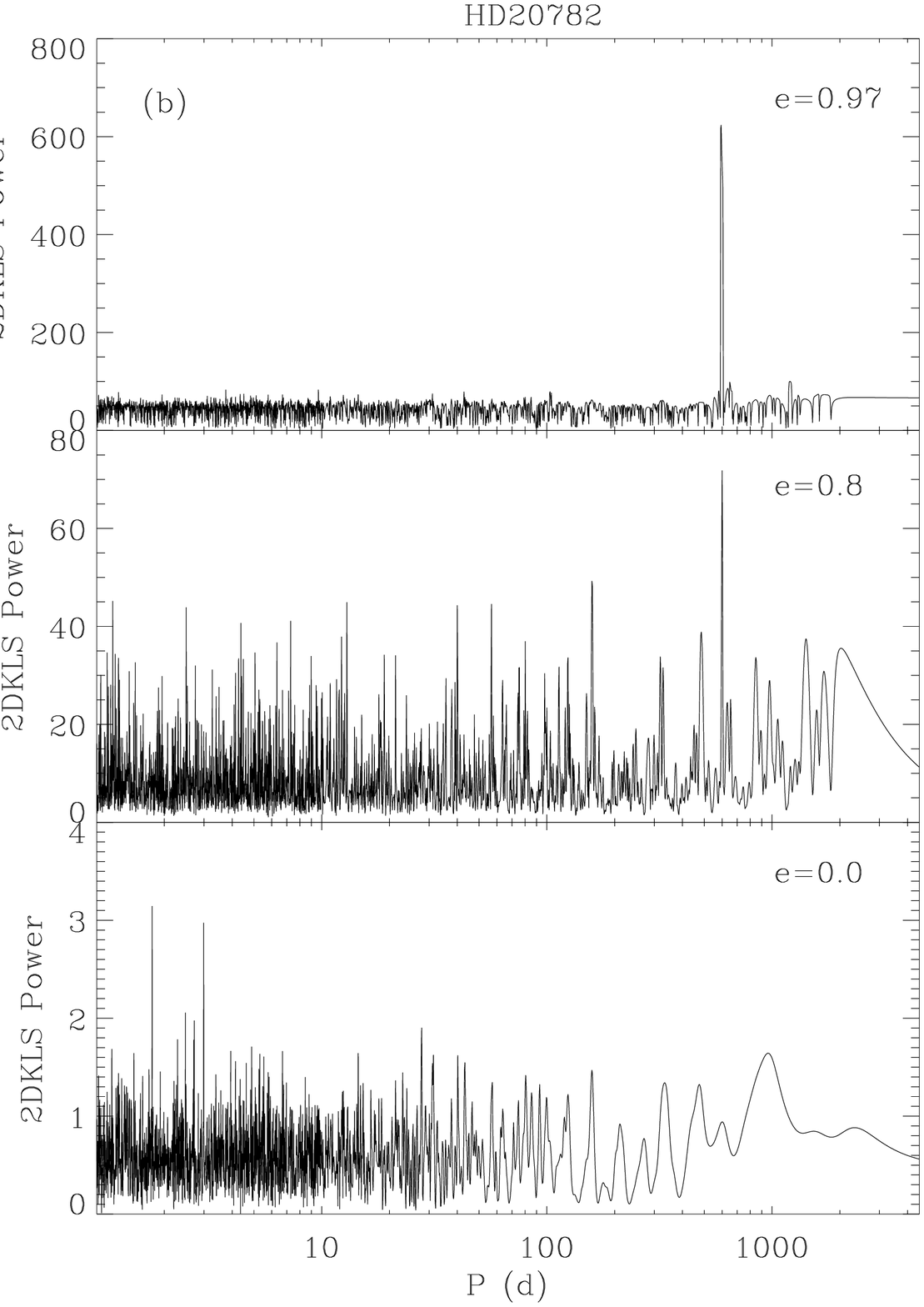,scale=0.47}
    \caption{(a) The 2D Keplerian Lomb-Scargle periodogram for
      HD20782. Dark areas indicate significant power. The dashed lines
      indicate the positions of the slices.
       (b) Three slices through the 2DKLS at the fit eccentricity
      (\emph{top panel}), $e=0.80$ (\emph{middle panel}) and $e=0.0$
      (\emph{bottom panel}). The signal at the fit eccentricity is
      very strong, barely detectable at $e=0.80$, and undetectable at
      $e=0.0$.}
    \label{fig:powercomp}
  \end{center}
\end{figure}

The 2DKLS has several advantages over the traditional LS
periodogram. First, it is sensitive to high-eccentricity planets,
which the traditional LS periodogram is not (as we show with an
example in Section \ref{sub:hd20782}). Second because the Keplerian
functional form fitted by the 2DKLS is a better representation of real
orbits, the peak in the 2DKLS power is higher than the traditional LS
power. Third, the width of the 2DKLS power peak more accurately
indicates the level at which the eccentricity is determined by a given
Doppler data set than the cross-terms in a single non-linear least
squares ``best'' Keplerian fit. (Real eccentricity uncertainties are
invariably much larger than the least-squares cross-term
uncertainties.) In addition, the 2DKLS can be used in a similar manner
to that of the CLEAN algorithm \citep{Hoegbom74}, if a simultaneous
fit of multiple Keplerians is also done. But perhaps most importantly
for our purposes, the 2DKLS allows for a simpler automation of the
planet detection process, as it much more rapidly narrows a Keplerian
trial fit on the ``correct'' best estimate of period and eccentricity.

\subsection{Application to HD\,20782}
\label{sub:hd20782}

The 2DKLS periodogram aids significantly in the detection of
high-eccentricity planets. As noted by \citet{JBT06} in their paper on
the extremely eccentric planet, HD\,20782b, detecting high
eccentricity planets using traditional periodogram methods is
extremely difficult. With the 2DKLS, however the detection of a planet
like HD\,20782b becomes straightforward.  The 2DKLS periodogram for
all AAPS data on HD\,20782 up to 2007, October 1 is shown in Figure
\ref{fig:powercomp}(a), along with slices through the 2DKLS at
$e=0.97$ (the best-fit Keplerian eccentricity), $e=0.8$ and $e=0.0$ in
Figure \ref{fig:powercomp}.  The planetary orbital signal is obvious
at $e$=0.97,  but the power peak becomes progressively smaller at
lower eccentricities; it is already difficult to discern at $e=0.8$,
and (most critically) completely undetectable at $e=0.0$. In other
words, this is a planet which an automated traditional LS
``first-pass'' analysis would never detect.

\begin{figure}[ht]
  \begin{center}
    \leavevmode
    \epsfig{file=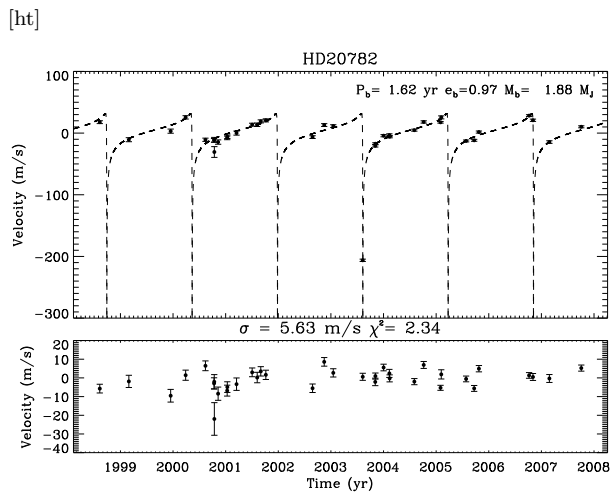,scale=0.47}
    \caption{Keplerian fit for HD\,20782, along with the residuals
      after subtraction of the best-fit model.}
    \label{fig:hd20782}
  \end{center}
\end{figure}

We update the \citeauthor{JBT06} parameters for this planet using the
most recent data in Table \ref{tab:hd20782} and show the revised fit
in Figure \ref{fig:hd20782}. The planet is in a highly eccentric
orbit, however, further refinement of the orbit with confirmation of
additional observations near periastron (i.e. near the large velocity
excursion) are important. The last excursion for the $e=0.97$ solution
occurred in the window 2008 June 18-21, when (unfortunately) HD\,20782
was inaccessibly behind the Sun, so further refinement of the orbit
will have to wait until the beginning of 2010. Measured velocities
are given in Table \ref{tab:vel20782}. To highlight the
importance of constraining eccentric planets in the narrow windows
when their velocities change most rapidly, we also show in Table
\ref{tab:hd20782} the results of the fit excluding the most extreme
velocity point. The significant change in the best-fit orbital
eccentricity (0.97 to 0.57)
that occurs as a result of removing just one data point highlights the
difficulties encountered in detecting and characterising eccentric
planets.

\begin{table}
\begin{center}
\caption{Updated Orbital Parameters of HD\,20782b}
\label{tab:hd20782}
\begin{tabular}{lcc}
\hline
Parameter & All obs. & without \\
 & & extreme pt. \\
\hline
$P$ (d)           & 591.9$\pm$2.8         & 577.9$\pm$2.6 \\
$K$ (m\,s$^{-1}$) & 185.3$\pm$49.7        & 21.7$\pm$1.2 \\
$e$               & 0.97$\pm$0.01         & 0.57$\pm$0.04 \\
$\omega$          & 147.7$\pm$6.5$^\circ$ & 98.6$\pm$5.7$^\circ$ \\
$T_0$ (JD-2451000)& 83.8$\pm$8.2          &175.9$\pm$8.4 \\
$M\sin i$ (\mjup) & 1.9$\pm$0.5           & 0.73$\pm$0.05 \\
$a$ (AU)       & 1.381$\pm$0.005 & 1.359$\pm$0.005 \\
N$_{obs}$      & 36 & 35  \\
RMS (m\,s$^{-1}$) & 5.6 & 4.8 \\
$\chi^2_\nu$      & 2.34 & 1.90 \\
jitter$^*$ (m\,s$^{-1}$) & 2.21 & 2.21 \\
\hline
\end{tabular}
\end{center}
$^*$Stellar jitter is calculated using the updated prescription of
J. Wright (2008, private communication).
\end{table}

\begin{table}
\caption{Velocities for HD\,20782 with corresponding measurement uncertainties.}
\label{tab:vel20782}
\begin{center}
\begin{tabular}{rr}
\hline
JD & RV \\
(-2451000)   &  (m\,s$^{-1}$) \\
\hline
  35.319456 &  17.7 $\pm$ 2.3 \\ 
 236.930648 & $-$10.7 $\pm$ 3.3 \\ 
 527.017315 &   3.2 $\pm$ 3.4 \\ 
 630.882407 &  25.5 $\pm$ 2.7 \\ 
 768.308854 & $-$10.8 $\pm$ 2.6 \\ 
 828.110660 & $-$11.8 $\pm$ 3.0 \\ 
 829.274491 & $-$10.8 $\pm$ 3.8 \\ 
 829.996250 & $-$30.6 $\pm$ 8.7 \\ 
 856.135301 & $-$14.5 $\pm$ 3.6 \\ 
 919.006597 &  $-$7.8 $\pm$ 2.9 \\ 
 919.996296 &  $-$5.8 $\pm$ 2.9 \\ 
 983.890093 &   0.000 $\pm$ 3.3 \\ 
1092.304375 &  13.7 $\pm$ 2.3 \\ 
1127.268137 &  13.5 $\pm$ 2.8 \\ 
1152.163079 &  19.0 $\pm$ 2.5 \\ 
1187.159653 &  20.7 $\pm$ 2.5 \\ 
1511.206498 &  $-$6.5 $\pm$ 2.3 \\ 
1592.048162 &  13.1 $\pm$ 2.3 \\ 
1654.960313 &  11.4 $\pm$ 2.2 \\ 
1859.305274 & $-$206.2 $\pm$ 1.9 \\ 
1946.138453 & $-$20.6 $\pm$ 2.0 \\ 
1947.122481 & $-$17.3 $\pm$ 1.6 \\ 
2004.001472 &  $-$4.2 $\pm$ 1.8 \\ 
2044.023669 &  $-$3.3 $\pm$ 2.2 \\ 
2045.960788 &  $-$5.8 $\pm$ 1.9 \\ 
2217.288060 &   4.7 $\pm$ 1.6 \\ 
2282.220295 &  18.1 $\pm$ 1.9 \\ 
2398.969109 &  17.5 $\pm$ 1.3 \\ 
2403.960670 &  25.5 $\pm$ 2.5 \\ 
2576.306902 & $-$12.7 $\pm$ 1.5 \\ 
2632.281289 & $-$11.7 $\pm$ 1.6 \\ 
2665.186505 &   1.7 $\pm$ 1.7 \\ 
3013.216410 &  28.3 $\pm$ 1.5 \\ 
3040.131498 &  20.8 $\pm$ 1.9 \\ 
3153.970057 & $-$14.5 $\pm$ 2.1 \\ 
3375.246543 &  10.2 $\pm$ 1.6 \\ 
\hline
\end{tabular}
\end{center}
\end{table}

\section{Simulations}
\label{sec:simul}

The goal of this work is to derive the underlying distributions of the
orbital parameters (period, eccentricity and minimum mass) as revealed
by our AAPS observations, so as to allow meaningful comparison with
planet formation and evolution models. Each object in our catalogue
has a different brightness, has different intrinsic velocity
variability, and has been observed at different epochs with varying
seeing and transparency conditions. The only way, therefore, to
understand the selection functions inherent to our data set is to
simulate it on a star-by-star and epoch-by-epoch basis. We have
therefore begun a major program of Monte Carlo simulations.

The time-stamps for the AAPS observations of our target stars were
used to create artificial data sets for single planets (modelled as a
single Keplerian using Equation \ref{eq:kepler} with an input period,
eccentricity and planet mass). The input periods on a logarithmic grid
from $\log_{10} P_i=0.0$ to 3.6 in steps of 0.3 (or 1 to 3981 days);
input eccentricities are on a grid from 0.0 to 0.9 in steps of 0.1;
and planet masses are on a grid with $M=(0.02, 0.05, 0.1, 0.2, 0.5,
1.0, 1.6, 2.3, 3.0, 4.0, 6.0, 9.0, 13.0,$ $20.0)$ in units of
$M_\mathrm{J}$. The semi-amplitudes for each artificial data set are
derived using the following:

\begin{equation}
K_i=\frac{M_i\sin i}{\sqrt{1-e_i^2}}\left [\frac{P_i(M_*+M_i\sin i)^2}{2\pi G}
\right ]^{-1/3} \label{eq:semiamp}
\end{equation}
where $M_*$ is the mass of the host star, $M_i$ is the planet's mass,
$i$ is the (unknown) inclination of the system and $G$ is the
gravitational constant. The subscript $i$ denotes the input
parameter. Measured parameters will be denoted with a subscript $m$.
Stellar isochrone masses from \citet{VF05} are used to estimate $M_*$
for each host star.

At each epoch, the ideal Keplerian has noise added -- which we model
at present as being Gaussian, with a width given by
the internal measurement uncertainty produced by that epoch's Doppler
analysis.  It is known that the measurement uncertainties themselves
do not follow a Gaussian distribution, for a variety of reasons. A more
realistic model for stellar noise in our simulations is currently
planned. This will incorporate stellar noise sources such
as magnetic activity \citep[e.g.][]{Wright05}, stellar oscillations
\citep{OTJ08} and stellar convection, and systematic measurement
effects.

The 2DKLS necessarily involves an increase in computation load
compared to the LS, meaning that parallelisation of code is
vital. Each simulation takes anywhere between 20 seconds and 10
minutes of CPU time per processor depending on the number of data
points. We have used the MPICH
implementation\footnote{http://www-unix.mcs.anl.gov/mpi/mpich1/index.htm}
of the \emph{Message Passing Interface} library to run our simulation
analysis system in parallel. The analysis of our early simulations
were run on a small Linux cluster comprising 10 processors at the
Anglo-Australian Observatory, as well as some of the 224 processors
available through the Miracle facility at University College,
London.\footnote{http://www.ucl.ac.uk/silva/research-computing/}
Subsequent analyses were moved to the Swinburne Centre for
Astrophysics and
Supercomputing\footnote{http://astronomy.swin.edu.au/supercomputing/}
in July 2007 utilising around 160 processors per star. 

One hundred simulations have been constructed for each set of
parameters ($P$, $e$, $K$), leading to 182\,000 simulations for each
target object. In this paper we focus on results for three stars:
HD\,20782, HD\,179949 and HD\,38382, whose relevant properties are
shown in Table \ref{tab:targets}. The first two objects each have a
known planet \citep{JBT06,TBM01}. However, to examine also the effects
of sampling and number of observations, we also consider a further two
subsets of the HD\,179949 data -- one using every second observation,
and the other every fourth -- and simulated those as well.

\begin{table}
\begin{center}
\caption{Properties of our target stars. $\Delta T$ is the time-span
  of the observations.}
\label{tab:targets}
\begin{tabular}{lccccc}
\hline
Star & V & Spec & $N$ & $\Delta T$ & median \\
 & (mag) & Type & & (days) & unc. (m\,s$^{-1}$) \\
\hline
HD\,20782$^*$  & 7.36 & G3V & 35 & 3119 & 2.27 \\
HD\,38382      & 6.34 & F8V & 17 & 2452 & 3.80 \\
HD\,179949$^*$ & 6.25 & F8V & 56 & 2626 & 5.26 \\
\hline
\end{tabular}
\end{center}
$^*$Indicates planet host star
\end{table}

\subsection{Distributions of fitted Keplerian parameters}
\label{sec:distro}

In a study of exoplanet parameter uncertainties, \citet{Ford05} found
that their distribution is typically non-Gaussian. In many ways this
is unsurprising, considering the correlations existing between
parameters, and as the description of Keplerian motion given by
Equation \ref{eq:kepler} is highly non-linear. 

\begin{figure}
  \begin{center}
    \leavevmode
    \psfig{file=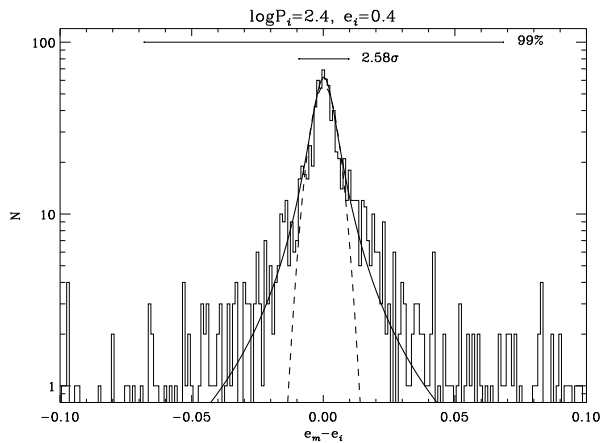,scale=0.47}
    \caption{Distribution of $e_m-e_i$ for HD\,179949 with
      $\log P_i=2.4$ and $e_i=0.4$. The dashed curve is a Gaussian fitted
      to the data, while the solid curve is a Lorentzian fit. 
      The 99\% confidence limit is 7.7 times larger than the
      equivalent 2.58$\sigma$ limit (were Gaussian
      statistics to apply).}
    \label{fig:lorecc}
  \end{center}
\end{figure}

\begin{figure}
  \begin{center}
    \leavevmode
    \epsfig{file=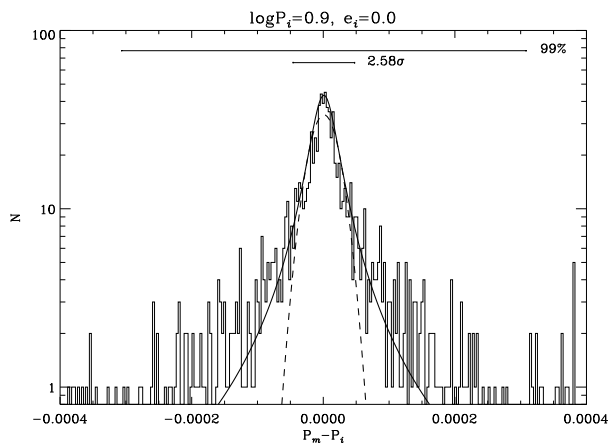,scale=0.47}
    \caption{Similar to Figure \ref{fig:lorecc} except for $P_m-P_i$ with
      $\log P_i=0.9$ and $e_i=0.0$. The 99\% confidence limit is 6.5 times larger than the
      equivalent 2.58$\sigma$ limit (were Gaussian
      statistics to apply). }
    \label{fig:lorper}
  \end{center}
\end{figure}

\begin{figure}
  \begin{center}
    \leavevmode
    \psfig{file=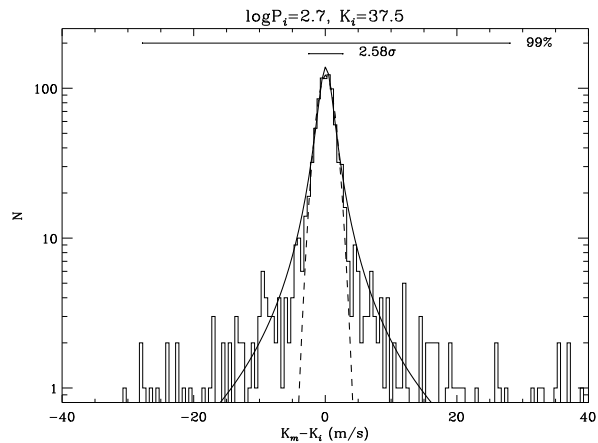,scale=0.47}
    \caption{Similar to Figure \ref{fig:lorecc} except for $K_m-K_i$
      with $\log P_i=2.7$ and $K_i=37.5$\,m\,s$^{-1}$. The 99\% confidence limit is 10.4 times larger than the
      equivalent 2.58$\sigma$ limit (were Gaussian
      statistics to apply).}
    \label{fig:loramp}
  \end{center}
\end{figure}

We can attempt to understand the uncertainty distributions of the
Keplerian orbital parameters that are produced in a least-squares fit
to an observed Doppler data set, by using our simulations to look at
the distributions of the orbital parameter {\em errors}\footnote{We
distinguish here between the {\em error} of a measurement (i.e. by how
much it is wrong, which can only be known when one knows the ``right
answer'' as in these simulations) and its {\em uncertainty} (i.e. an
estimate of the quality of a measurement in the absence of knowing the
``right answer'').} (i.e. the differences between the input ``$i$''
and measured ``$m$'' simulation values). Figure \ref{fig:lorecc} shows
the distribution of the $errors$ in eccentricity for HD\,179949, with
$\log P_i$ fixed at 2.4 and $e_i$ fixed at 0.4 (i.e. including
eccentricity errors at all planet masses). The dashed line is a
Gaussian fit to the histogram, while the solid line is a Lorentzian
fit. The wings of the distribution diverge significantly from a
Gaussian and are better matched by the Lorentzian. Note that we are
not claiming the distribution \emph{is} Lorentzian, simply that the
extended wings of this function is a better model of the extended
wings of the observed distribution.  Similarly, the distribution of
measured periods and semi-amplitudes, shown in Figures \ref{fig:lorper}
and \ref{fig:loramp} respectively, are non-Gaussian, with extended
Lorentzian-like wings. The special case of $e_i=0$ is not only
non-Gaussian, but also non-symmetric.  An obvious consequence of this
and the analysis of \citet{Ford05} is that Gaussian statistics for the
orbital parameter estimates (e.g. 3$\sigma$ or 5$\sigma$ limits)
cannot be used as criteria for exoplanet detection. 

To demonstrate this, we show in Figures
\ref{fig:lorecc}-\ref{fig:loramp} both the 99\% limits of these error
distributions {\em and} the data ranges that would correspond to such
a limit were Gaussian statistics to apply (i.e. 2.58$\sigma$). In
every case the {\em observed} 99\% confidence limits are much larger
by factors of $\sim$5-10. Thus the {\em real} uncertainty on a
Keplerian fit parameter is significantly larger than that one would
predict based on Gaussian statistics alone,  and  Gaussian statistics
must be either avoided, or suitably modified, in any set of exoplanet
detection criteria. We discuss a set of criteria that take this effect
into account in Section \ref{sec:detect}, and in particular, we
examine empirical relationships which can be used to calibrate and
derive robust confidence limits for the orbital elements produced by
least-squares Keplerian fits in Section \ref{sub:errors}.

\section{Detection criteria}
\label{sec:detect}

One of the important practical considerations of our simulation system
is that it must be automated. We therefore require a set of criteria
to decide whether a planet has been detected, without any human
intervention. These will be applied to the results of Keplerian fits
to simulated data that should be both robust and not add significantly
to the time budget of our analyses.

When determining adequate criteria, there are two important differences
between the analysis of real and simulated data that must be considered.
First, there is a difference between simply trying to determine whether
one of 250 target stars has a planet, and whether one of
millions of data sets has a real planet or not. In the former case, as
much time can be spent as needed on trying to decide ``by eye''
whether it is real or not. For latter case automation is essential.

Second, the aim of these simulations is not the same as that for
planet discovery. In the latter, the bias is towards seeing whether a
planet has been found orbiting a target star, with subsequent
observations being used to confirm or deny its status. For a simulated
detection there are no subsequent observations -- one has to decide
the status using \emph{only} the simulated data. Moreover, there is a
simulated planet present in \emph{every} data set. What we
need to know whether it has been detected with sufficient robustness
that we can be sure (within a given confidence level), that
it is a real detection. As such the
simulated detection criteria will almost always be more stringent than
the criteria used for the discovery of an exoplanet from actual planet
search data.

Considered another way, these simulations are aiming to generalise the same
process that is used in estimating the $1/V_{\mathrm{max}}$ volume
that is represented by each star going into the estimation of a
luminosity function. In this case we are estimating for each star
$V_{\mathrm{max}}$ in the ($P$, $e$, $K$) phase space that is
accessible to a given set of data. This means that a selection of a
set of detection criteria that determines $V_{\mathrm{max}}$ is
arbitrary -- it determines the sensitivity, but not the results, of
our survey. Have too loose a set of detection criteria, and you find
lots of objects, and have a large $V_{\mathrm{max}}$, but are subject
to false positives. Have too tight a set of detection criteria, and
you find few objects and have a small $V_{\mathrm{max}}$, but are much
less subject to the biases associated with false positives.

\subsection{Previous sets of  detection criteria}
\label{sub:other}

There are several methods that have previously been used to determine
the reality of a planet detection. \citet{MBV05} presented an
excellent discussion on two different approaches to the False Alarm
Probability or FAP. The FAP is the probability that the best-fit model
Keplerian could have arisen simply as a result of noise
fluctuations. The first involves the classical $F$-test
\citep{Beving69} which has several weaknesses: it assumes that the
uncertainties of the measurements follow a Gaussian distribution --
even the smallest deviation from normality has been reported to be
extremely non-robust \citep{Lindman74}; it cannot properly account for
the actual uneven temporal sampling of the observations
\citep[e.g.][]{MBV05}; and it depends on the number of independent
frequencies -- a number which can only be approximated. Both
\citet{Cumming04} and \citet{CBM08} used the $F$-test to investigate
the detectability of planets in Doppler surveys based on the
analytical FAPs.

The other method presented by \citet{MBV05} is empirical and involves
creating 1000 or more quasi-artificial data sets by generating
randomly scrambling the velocities, but keeping the times the same, and
then analysing the new sets in the same way as the original data. The
number of $\chi^2_\nu$ values similar to the value for the candidate
detection is then used to construct a FAP. This approach has the
advantage that the distribution of uncertainties and temporal sampling
of the observations are unimportant. Used alone it cannot distinguish
between peaks with similar significance in a power spectrum of the
actual observations, as these are likely to have similar FAPs.

Because of the large number of simulations we plan to carry out, it is
important we have a simple set of criteria that can quickly test the
reliability of a detection. This automatically rules out several
approaches that are in themselves computationally intensive; in
particular the \citeauthor{MBV05} scrambling method to determine a FAP
described above would add considerably to the time budget of our
simulation analysis. The $F$-test used by \citet{Cumming04} is also
inappropriate for two reasons: first, AAPS data has uneven temporal
sampling, and second, because we have incorporated our velocity
measurement uncertainties -- which do not follow a Gaussian
distribution -- into the our noise estimates, the simulated velocities
show small departures from a pure Gaussian. We have therefore developed our
own set of criteria, discussed below.

\subsection{Our criteria}
\label{sub:3pt}

There are two final points to consider before we present our
criteria. First, the criteria we use must be ``blind'' -- that
is, they must only be based on measured quantities, and have no
reliance on the input orbital parameters of the simulations. Second,
the number of false positives should be as low as
possible. The following are by no means
the only criteria that we could use, however, as we show in Section
\ref{sub:false}, they produce an acceptable fraction of false
positives. The set of criteria we have found to be useful are:

\renewcommand{\theequation}{\roman{equation}}
\setcounter{equation}{0}
\begin{equation}
\;\;\;\;\;\;\;\;\;\;\mathrm{RMS(sim)}\ge\mathrm{RMS(res)} \label{eq:rms}
\end{equation}
\emph{The RMS of the simulated observations must be greater than or
  equal to the RMS of the residuals of the best-fit model.} That is,
by subtracting a Keplerian model from the time series, the overall
scatter should decrease, rather than increase.

\begin{equation}
\;\;\;\;\;\;\;\;\;\;K_m \ge 2\delta K+\mathrm{RMS(res)} \label{eq:2dKrms}
\end{equation}
\emph{The measured semi-amplitude must be greater than or equal to
  twice the semi-amplitude uncertainty, plus the RMS of the residuals of the
  best fit model.} The uncertainties from the fit procedure we have used are
  correlated -- the off-axis terms in the covariance matrix calculated
  in the non-linear least squares fit are non-zero -- which means that
  the fit uncertainties are lower limits. We approximate the semi-amplitude
  uncertainties here as twice the fit uncertainty plus the RMS of the
  residuals of the best-fit model. This criterion rejects poorly
  constrained semi-amplitudes (and therefore poorly-constrained planet
  masses).

\begin{equation}
\;\;\;\;\;\;\;\;\;\;P_m \ge 2\delta P \label{eq:p2perr}
\end{equation}
\emph{The measured period must be greater than or equal to twice the
  period uncertainty}. As with the semi-amplitude uncertainties in
criterion (\ref{eq:2dKrms}) above, the period uncertainties are
underestimated. We double the fit uncertainty to reject poorly
constrained periods.

\begin{equation}
\;\;\;\;\;\;\;\;\;\;\chi^2_\nu\le 3 \label{pt:chisq}
\end{equation}
\emph{The $\chi^2_\nu$ value must be less than or equal to 3.} This is a
  somewhat arbitrary cut, however it significantly reduces the number
  of false positives at high eccentricities, as discussed in Section
  \ref{sub:false}.

As a simple test of our criteria we have used them to check whether
each of the published planets in the AAPS target catalogue would be
found. As part of this test we include a stellar jitter term in our
fits, despite not including it in our simulations. This is because
jitter is present in the real observations (and contributes to the
individual measurement uncertainties of those real observations),
while it is not present in our simulated observations, which have only
had Gaussian noise added to them. By including the appropriate jitter
values in our analyses (J. Wright, 2008 private communication), we
find that 15 of the known planets satisfy our criteria. Excluding the
$\chi^2_\nu$ cut, all bar one of the planets pass the test. This bares
no reflection of the reality of the planets, but rather a reflection
of the strictness of our criteria.  Other effects at play here are the
presence of multiple companions, and accuracy of the jitter
measurements used, which is only around $\pm$\,50\% (J. Wright,
private communication; note that the jitter contains unquantified
time-variability).

\subsection{Orbital parameter confidence limits}
\label{sub:errors}

\begin{figure}
  \begin{center}
    \leavevmode
    \psfig{file=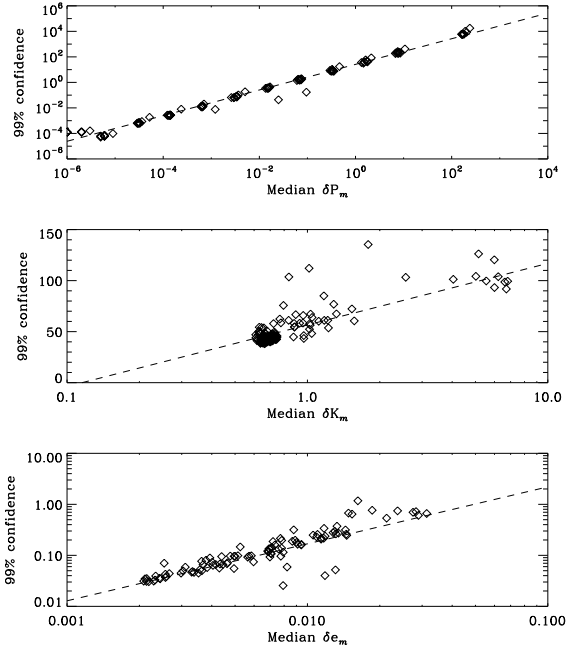,scale=0.47}
    \caption{Plot of 99\% limits in errors (i.e. difference between input and
      measured orbital parameters)  versus median
      uncertainty in the parameter from least-squares Keplerian fits to
      the simulated data, for each orbital parameter for HD\,20782. The
      dashed line shows a power law with the parameters listed in
      Table \ref{tab:errors}.}
    \label{fig:errors}
  \end{center}
\end{figure}

\begin{table}
\begin{center}
\caption{Power law exponents for each star. The power laws have the form
10$^\alpha X^\beta$, for $X=\delta P_m, \delta e_m$ and
$\alpha+\beta\log X$ for $X=\delta K_m$.}
\label{tab:errors}
\begin{tabular}{cccc}
\hline
Parameter  & HD\,20782     & HD\,38382      & HD\,179949     \\
\hline
$\alpha_P$ & 1.41$\pm$0.03 & 1.67$\pm$0.02  & 1.50$\pm$0.02  \\
$\beta_P$  & 1.00$\pm$0.01 & 1.07$\pm$0.01  & 1.02$\pm$0.01  \\
\\
$\alpha_K$ & 56.6$\pm$1.3  & 90.0$\pm$6.0   & 75.6$\pm$4.3    \\
$\beta_K$  & 60.3$\pm$4.9  & 134.5$\pm$15.6 & 155.0$\pm$10.2  \\
\\
$\alpha_e$ & 1.46$\pm$0.13 & 1.82$\pm$0.26  & 2.10$\pm$0.26  \\
$\beta_e$  & 1.12$\pm$0.06 & 1.33$\pm$0.15  & 1.41$\pm$0.13  \\
\hline
\end{tabular}
\end{center}
\end{table}

We have seen in the discussion of distribution functions in Section
\ref{sec:distro} that Gaussian statistics cannot be used to model the
distribution of errors in orbital parameters that arise from
least-squares Keplerian fits to our simulated data (and obviously they
similarly cannot be used to model the uncertainties in orbital
parameters for detected Doppler planets from real data sets either).

The orbital parameter error distribution functions \emph{do} contain
information that is useful, however, in that they allow us to
empirically calibrate the relationship between the uncertainties that
arise from the covariance matrix in a least-squares Keplerian fit
(i.e. the source of the traditional uncertainties in orbital
parameters produced in analysing Doppler data sets) and our observed
error distributions. To examine these relationships we compare the
uncertainties from the least-squares Keplerian fit and the 99\%
confidence range from the simulations.

In Figure \ref{fig:errors} we show the 99\% confidence limit as a
function of the median uncertainty of the fit $\delta P_m$ (top),
$\delta K_m$ (middle) and $\delta e_m$ (bottom) for HD\,20782 at at
fixed pairs of $P_i$ and $e_i$ or $P_i$ and $K_i$ (as in Section
\ref{sec:distro}). There is clearly a correlation between these
parameters, which we have characterised in the figure with a power-law
for each parameter.  The power laws have the form: 99\% confidence =
10$^\alpha X^\beta$, where $X=\delta P_m, \delta e_m$ and 99\%
confidence = $\alpha+\beta\log\delta K_m$ for semi-amplitude. The
parameters $\alpha$ and $\beta$ from these fits for HD\,20782 are
listed in Table \ref{tab:errors} (along with the equivalent parameters
for similar fits to the equivalent data for HD\,38382 and HD\,179949).

For the period data, it is clear that the power-law slope is consistent with
an index of 1 -- that is, the 99\% confidence limits are linearly related
to the 1$\sigma$ fit uncertainties by factors of 26-47, or equivalently
Gaussian statistics overestimate the 99\% limits for a period determination
from these simulated Doppler data by factors of between 10-18.
For eccentricity the correlation is weaker, and the power-law fit indicated
is slightly above 1. More importantly, the 99\% confidence limits are
about 10-50 times larger than those which would be derived from simply
trusting the Gaussian nature of the uncertainties coming from 
a least-squares Keplerian fit. This reflects the fact that eccentricity is,
in general, very poorly constrained by Doppler data sets, as was seen in our
analysis of the 2DKLS.
The correlation between the observed 99\% confidence limits
and the Keplerian fit uncertainties for semi-amplitude $K$ is poor,
and again the 99\% confidence limits in $K$ are larger than those
one would predict from the Keplerian fit uncertainties.

From our analysis of these three data sets, it would not appear that
there are any general conclusions that can be reached, for every star
and every data set, on how to relate {\em real} 99\% confidence limits
to Keplerian fit uncertainties (other than that Keplerian fit
uncertainties significantly under-estimate -- by factors of greater
than 10 -- the real confidence limits). However, it is clear that for
a {\em given} simulated data set, that meaningful correlations can be
derived and applied. We have therefore used these relationships to
convert fit uncertainties into meaningful confidence limits for our
subsequent analysis of false positives.

\subsection{False positives}
\label{sub:false}

It is important that the number of false positives (i.e. the
number of incorrect detections) triggered by our detection
criteria be (a) quantifiable, and (b) as small as possible. 
We adopt as our ``incorrectness'' criterion that the measured
orbital parameter differs from the input orbital parameter by more than
the 99\% confidence limit for that orbital parameter (as derived in 
Section \ref{sub:errors}).

For each simulation that results in a detection, we
test for ``correctness'' by asking; 
\begin{enumerate}
\item is the error in period (i.e.
the difference between measured period and input period) less than the
99\% confidence limit (as derived from calibrating the least-squares fit
period uncertainty to a true confidence limit as described in
\ref{sub:errors})? If the error is larger than the 99\% limit, we call
the period incorrect. We also ask,
\item is the error in the the semi-amplitude larger than the 99\% confidence
limit? If it is then we call the semi-amplitude incorrect.
\end{enumerate}

If both the period and the semi-amplitude are determined to be incorrect (at the
99\% level), we describe this as an incorrect detection, or false positive. The false positive
rate is then the ratio of the number of incorrect detections to the total number
of detections. Averaging over all parameters, we find the false
positive rate due to incorrect period and semi-amplitude to be 1.2\%
for HD\,179949, 2.2\% for HD\,20782, and 9.0\% for HD\,38382. In the
sections that follow, we look in more detail at these numbers and at
their trends as a function of input orbital parameters.

\subsubsection{The $\chi_{\nu}^2\le 3$ Detection Criterion }
\label{text:chicut}

\begin{figure}
  \begin{center}
    \leavevmode
    \psfig{file=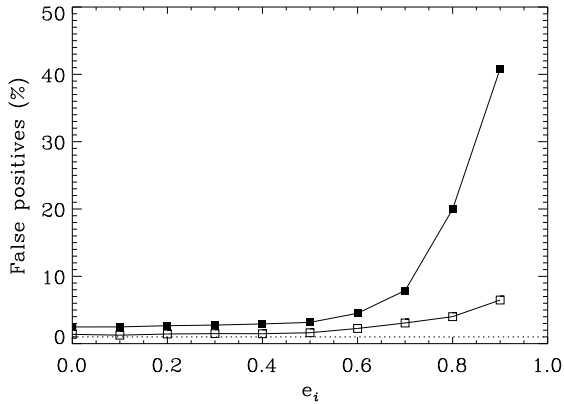,scale=0.47}
    \caption{False positives (open squares) for HD\,20782 as a 
      function of simulated input eccentricity. The
      filled squares represent the corresponding false positives
      excluding the $\chi^2_\nu$ detection criterion (see section \ref{text:chicut}
      in the text). Without this  $\chi^2_\nu$ cut, the simulations would
      be subject to an unacceptable level of false positives at
      high eccentricities.}
    \label{fig:testfalse}
  \end{center}
\end{figure}

We are now in a position to demonstrate our reasons for include this particular
criterion, which we do in Figure \ref{fig:testfalse}, which shows the false
positive fraction as a function of simulated input eccentricity both
with this criterion applied (open squares) and not applied (filled squares).
It immediately becomes apparent that without this particular criterion being
applied, our data set is subject to an unacceptably large fraction of false 
positives at high eccentricity. Even with this criterion being applied
the number of false positives shows an increase over the ``base'' level of
1-2\% seen at low eccentricity, up to 6\% at $e$=0.9. (Similar patterns are
seen for the other sets of simulations for the other stars.)

\subsubsection{Eccentricity}

\begin{figure}
  \begin{center}
    \leavevmode
    \psfig{file=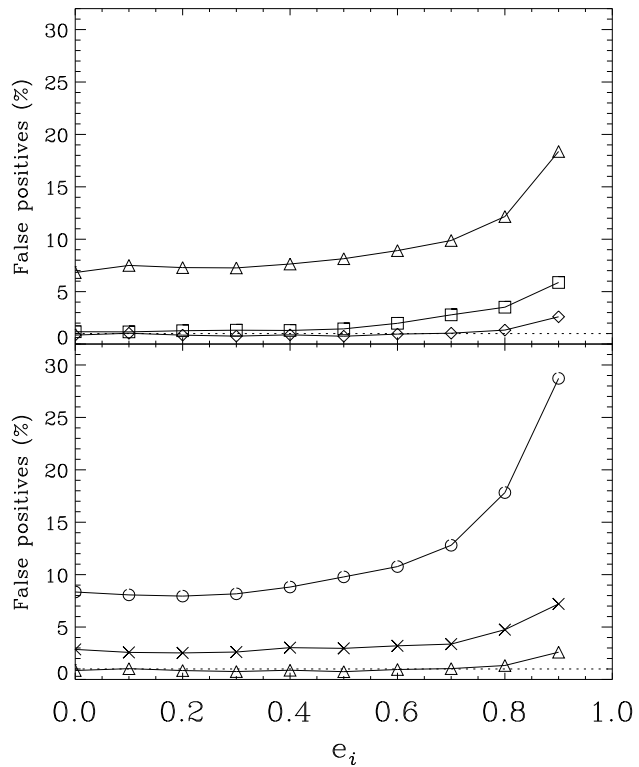,scale=0.7}
    \caption{\emph{Top panel}: False positives for HD\,179949
(diamonds), HD\,38382 (triangles) and HD\,20782 (squares) as a
function of input eccentricity. \emph{Bottom panel}: False positives
of each of the HD\,179949 subsets -- the full set of 56 epochs
(triangles), the 28 epoch subset (crosses) and the 14 epoch subset
(circles) -- demonstrating the significant increase in false positive
detections at low observation density. A dotted line is shown at the
1\% false positive level.}
    \label{fig:efp}
  \end{center}
\end{figure}

We show in the top panel of Figure \ref{fig:efp} the rate of false
positives for each object as a function of input eccentricity,
at all values of $P_i$ and $M_i$. For HD\,179949 and HD\,20782, the
percentage of false positives remains at $\sim$\,1\% up to $e_i\approx
0.6$ and then increases to 3-6\% at $e_i=0.9$. In the case of
HD\,38382, the false positive rate is around 7\% up to $e_i\approx
0.5$, increasing to 18\% at $e_i=0.9$. The higher false positive rate
for this star is due to its having much fewer observations (just 17
epochs) than the other two stars --  fewer observations make it harder
to detected an exoplanet, and conversely makes that data set more
subject to false positive detections.  To demonstrate this, we show in
the bottom panel of the same figure the false positive rate for the
three HD\,179949 subsets. The 28 epoch subset (crosses) has 2-3  times
as many false positives as the full HD\,179949 data set (with 56
epochs; diamonds)  and the 14 epoch subset matches the HD\,38382 false
positive rate at low eccentricities and then becomes worse as
eccentricity increases. The star-by-star approach in this case
reproduces the expected behaviour -- more observations give more
confidence in an exoplanetary detection.

\begin{figure}
  \begin{center}
    \leavevmode
    \psfig{file=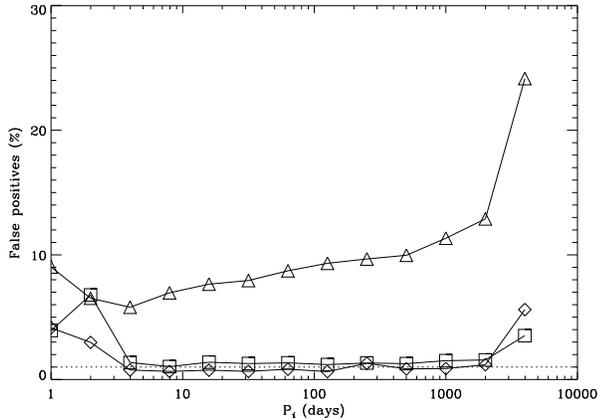,scale=0.47}
    \caption{False positives of the measured period values for
      HD\,179949, HD\,38382 and HD\,20782 as a function of input
      period. Symbols have the same meaning as Figure \ref{fig:efp}.}
    \label{fig:pfp}
  \end{center}
\end{figure}

\begin{figure}
  \begin{center}
    \leavevmode
    \psfig{file=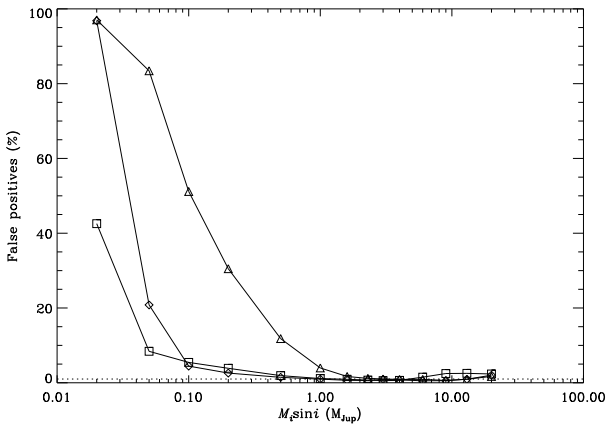,scale=0.47}
    \caption{False positives of the measured
      semi-amplitude values for HD\,179949, HD\,20782 and HD\,38382 as
      a function of input planet mass. Symbols have the same meaning
      as Figure \ref{fig:efp}.}
    \label{fig:mfp}
  \end{center}
\end{figure}

\begin{figure}
  \begin{center}
    \leavevmode
    \psfig{file=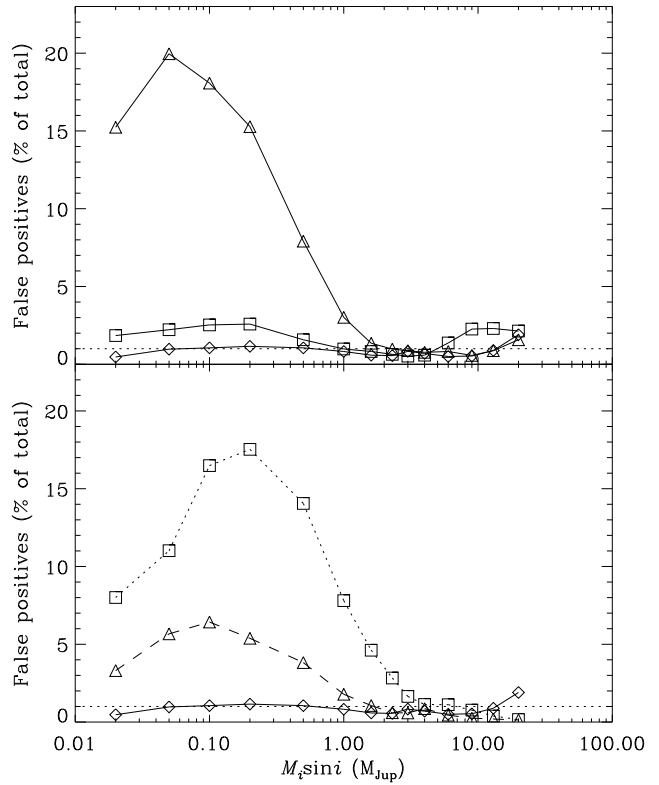,scale=0.7}
    \caption{False positives as a function of total number of
      simulations for HD\,179949 (diamonds), HD\,20782 (squares) and
      HD\,38382 (triangles) (\emph{top panel}) and for the HD\,179949
      subsets (\emph{bottom panel}): $N=14$ -- dotted line; $N=28$ --
      dashed line; $N=56$ -- solid line.}
    \label{fig:correct_all}
  \end{center}
\end{figure}

\subsubsection{Period}

The false positive rates for our three stars are shown as a  function
of $P_i$ (at all values of $e_i$ and $M_i$) in Figure \ref{fig:pfp}.
At periods longer than the time-span of the observations and below
$\sim$\,4\,d, the rate of false positives is 4-9\% for HD\,179949 and
HD\,20782. As with eccentricity, the false positive rate is below
$\sim$\,2\% for these stars. For HD\,38382 the false positive rate
steadily increases as a function of $\log P$ from around 6\% at short
periods, to $\sim 25$\% at longer periods -- again this is due to the
sparser sampling of the HD\,38382 data.

\subsubsection{Planet mass}

Finally, the false positive rate as a function of $M_i$ (at all values
of $P_i$ and $e_i$) is shown in Figure
\ref{fig:mfp}. While the large numbers of false positives at low mass
might at first appear troubling, it must be remembered that there is
a significant selection effect {\em against} finding objects at very
low masses. Therefore the number of correct detections will decline
steeply at very low masses (as low mass planets are very hard to
detect), while the incorrect detection rate should remain roughly
constant. This will lead to a systematic increase in the false
positive rate at very low masses. 

We test this idea by showing the false positives as a function of the
total number of simulations for each star in the top panel of Figure
\ref{fig:correct_all}. The \emph{incorrect} detections seem to be
approximately constant for HD\,179949 and HD\,20782, with values of
$\sim$1-2\% for the former and $\sim$2-3\% for the latter. In the case
of HD\,38382, however, there is an increase from about
1.6\,$M_{\mathrm{Jup}}$ up to around 20\%. Comparing HD\,38382 to the
three subsets of HD\,179949 (bottom panel of Figure
\ref{fig:correct_all}), we see the cause of the increase is
observation density. The subset with $N=14$ -- approximately the same
as HD\,38382 -- suffers from the same effect. The $N=28$ subset shows
the effect to a lesser degree and it is almost completely removed in
the $N=56$ subset. This shows one weakness in our detection criteria
and we are currently investigating ways to minimise the problem.

\begin{figure*}
  \begin{center}
    \leavevmode
    \psfig{file=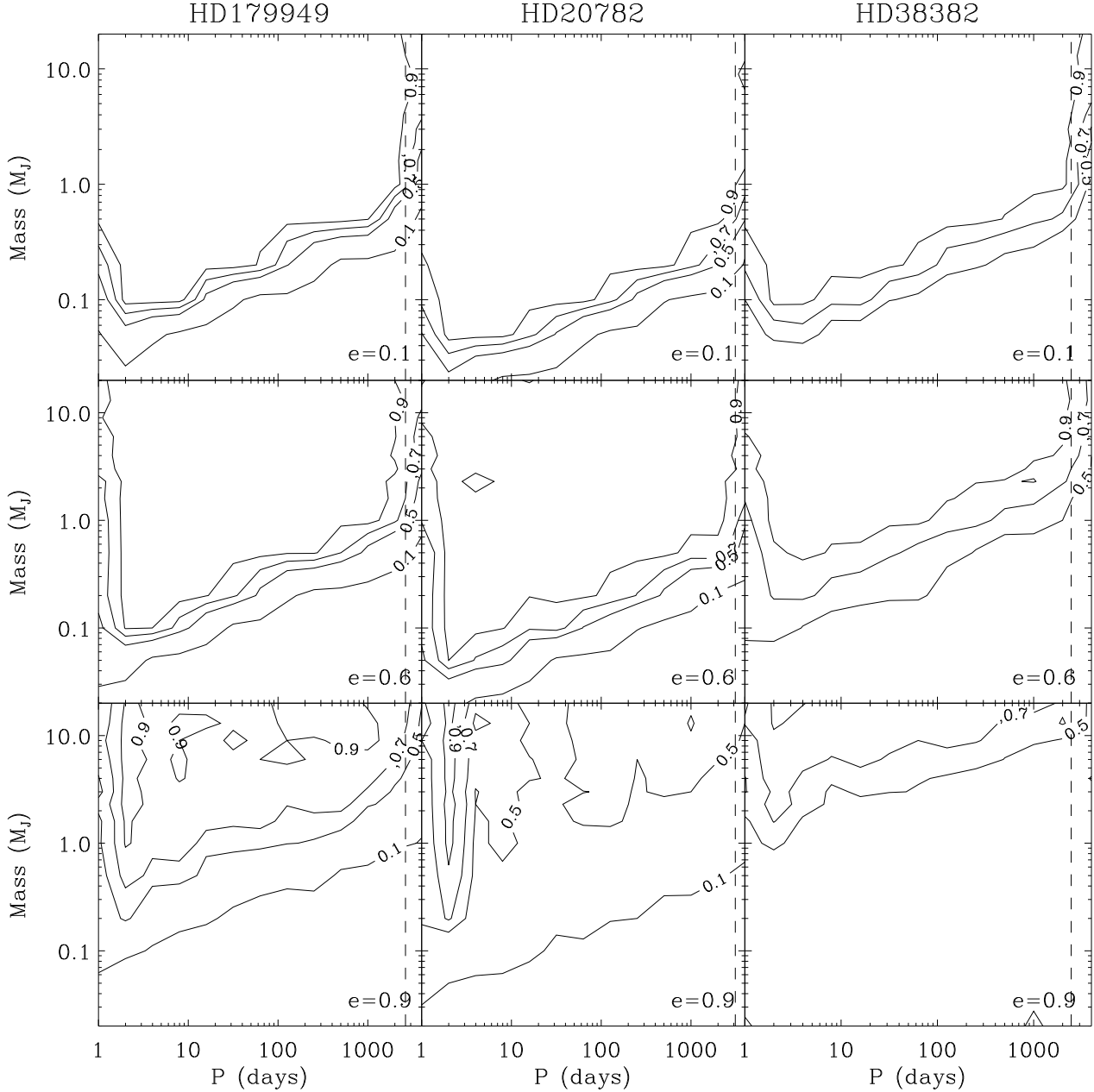,scale=0.9}
    \caption{Detectability of planets as a function of input period
      and input planet mass at 3
      different values of eccentricity (0.1, 0.6, 0.9) for HD\,179949
      (56 epochs, $\Delta T=2626$\,d), HD\,20782 (35 epochs, $\Delta
      T=3119$\,d) and HD\,38382 (17 epochs,
      $\Delta T=2452$\,d). False positives have been subtracted from
      each star. The vertical dashed line in each panel
      indicates $\Delta T$.}
    \label{fig:Pselfunc1}
  \end{center}
\end{figure*}

\section{Results}
\label{sec:prelim}

Our automated detection criteria, and a means to analyse
false positive rates, place us in a position to examine in
detail the observational biases present in our data for the
three stars under simulation. We define the detectability \dpem\ as
the detection rate as a function of
$P_i$, $e_i$, and $M_i$ -- in the case of our simulations 
we performed 100 realisations at each point in ($P_i$,$e_i$,$M_i$) space,
so we divide the number of detections by 100. 
False positives have been removed from the \dpem\ for each set of
simulations. In Figure \ref{fig:Pselfunc1}, we show
contour maps of detectability, for each
star, as a function of input period and planet
mass for three different eccentricities ($e=0.1$, $0.6$, $0.9$).
Note that the HD\,179949 map is for the full set of 56 observations. The
vertical dashed line indicates the time-span of the observations ($\Delta T$).
These detectability surfaces display a number of common features.

First, it can be clearly seen that detectability is quite low
at periods longer than the time-span of the observations 
(as seen by both \citet{Cumming04} and \citet{WEC06}), and
is also low at short periods (below $\sim$2\,d).
Both these effects are precisely what one would naively expect,
based on our ability to sample the Doppler variability of our target
stars. And both are reflected in the properties of the currently
detected exoplanets. Few exoplanets are known at periods of
10 years or longer as a result of Doppler searches which are
based on around a decade's worth of data. And, no Doppler 
exoplanet has been found at periods of less than $\sim$2 days
without first being detected via a transit event. 

Second, \dpem\ as a function of planet mass, for a given eccentricity,
reveals the same general pattern for each star, with detectability
decreasing with increasing period. This is
not surprising, since (to first order) for data sets with similar Doppler
measurement precision over time, the ability to detect an exoplanet is determined
by the size of the semi-amplitude $K_m$ relative to that precision, which
is in turn (from Equation \ref{eq:semiamp}) a function of $P_m$ of the form
$$M \propto 1/\sqrt[3]{P}.$$

The combination of these two effects means that the general form of
the \dpem\ surface is one of a transition region (with slope $M
\propto 1/\sqrt[3]{P}$) dividing highly detectable planets (or
generally higher mass and shorter period) from undetectable ones (or
lower mass and longer period), modified by a cut-on at short periods
of $\sim$ 2 days (determined by the shortest data sampling obtained),
and a cut-off at long periods (determined by the length of the
observation string, ${\Delta}T$). These are the general behaviours that
one would expect. Of more interest is the detailed behaviour these
surfaces reveal for each target.

In particular, for example, we see that the steepness of the
``transition region'' between highly detectable, and mostly
undetectable planets is a strong function of the number of
observations obtained. The slope in the transition region is steep for
HD\,179949 and HD\,20782, but shallow for the more poorly sampled
HD\,38382 data.

Moreover, the slope of the transition region is also a function of
eccentricity, for all three stars, highly eccentric planets have a
gently sloping transition region that mostly fills the entire
available mass-period plane. Indeed, it is only at short periods
that even the best sampled data sets have high detectability.

To assist in the visualisation of a more detailed analysis of these
general trends, we define the integrated detectability,
$D^\prime_{\mathrm{int}}$, such that $D^\prime_{\mathrm{int}}(P_i)$ is
simply the detectability at a given period $P_i$, over \emph{all}
$e_i$ and $M_i$ (and by extension $D^\prime_{\mathrm{int}}(M_i)$ and
$D^\prime_{\mathrm{int}}(e_i)$ are defined as the detectability
over all the other relevant parameters in each case). In particular,
we pay special attention to the three subsets of the HD\,179949 data
set (as described in Section \ref{sec:simul}) in order to examine the
impact of data sampling.

\subsection{Eccentricity}
\label{sub:ecc}

\begin{figure}
  \begin{center}
    \leavevmode
    \epsfig{file=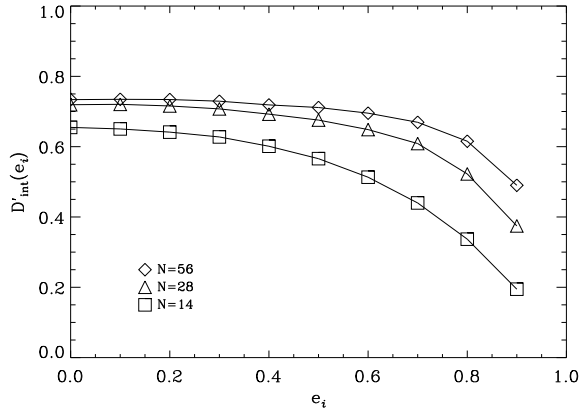,scale=0.47}
    \caption{Fraction of simulated planets re-detected as a function
      of input eccentricity, or $D^\prime_{\mathrm{int}}(e_i)$, for
      HD\,179949 with $N=14, 28, \&\ 56$ with false positives
      removed. This is over all
      periods and semi-amplitudes and shows the importance of data
      sampling and number of epochs for detecting highly eccentric
      planets. The points are connected to identify different trends
      for each star.}
    \label{fig:eselfunc2}
  \end{center}
\end{figure}

Figure \ref{fig:eselfunc2} shows the integrated detectability for each
of the HD\,179949 subsets as a function of eccentricity,
$D^\prime_{\mathrm{int}}(e_i)$; recall that false positives have been
removed. There is a clear difference at high eccentricity between each
of the subsamples. At N=14 (squares), $D^\prime_{\mathrm{int}}(e_i)$
drops significantly at $e_i=0.5$ and higher. At higher values of N (28
-- triangles; 56 -- diamonds), the drop-off starts at higher
eccentricity ($e_i>0.6$ for N=28 and $e_i>0.7$ for N=56). Below each
of these values, $D^\prime_{\mathrm{int}}(e_i)$ is approximately
constant. As demonstrated elsewhere \citep[e.g.\ ][]{Cumming04}, the
implication here is that higher observation density makes detection of
high eccentricity planets more likely.

\begin{figure}
  \begin{center}
    \leavevmode
    \epsfig{file=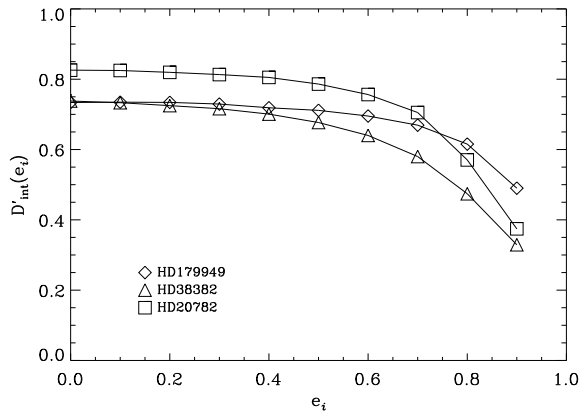,scale=0.47}
    \caption{Similar to Figure \ref{fig:eselfunc2}, except for
      HD\,20782, HD\,38382 and HD\,179949. }
    \label{fig:eselfunc1}
  \end{center}
\end{figure}

The subsets of HD\,179949 show significant variations in
$D^\prime_{\mathrm{int}}(e_i)$, but is there a difference between stars? Figure
\ref{fig:eselfunc1} shows $D^\prime_{\mathrm{int}}(e_i)$ for each of the three
objects, HD\,179949, HD\,20782 and HD\,38382. The shape of the curves,
while in general decreasing at higher eccentricities, is different for
each object. For example, when $e_i\le 0.1$ HD\,179949 has the lowest
fraction of planets redetected, however when $e_i\ge 0.8$ it has the
highest.
Thus data sampling and quality are fundamental to the selection
effects present in planet search observations and  a simple
parametrisation of the detectability of exoplanet parameters using
``whole-of-survey'' metrics -- e.g. \citet{Cumming04} -- \emph{cannot
be done}. As an example, consider the case of HD\,179949. Of the three
sets of observations we discuss here, the HD\,179949 data have the
highest median measurement uncertainty (5.26\,m\,s$^{-1}$), and one
might naively expect it's detectabilities to be the lowest. However,
the observation density (equal to the observation time-span/number of
observations or $\Delta T/N$) is the highest at 47\,days/epoch, which
should counteract the first effect to some degree. It is not
intuitively clear how to parametrise and compare the detectability of
the HD\,179949 observations, with, for example, that of the HD\,38382
observations -- which have lower observation density but also lower
median measurement uncertainty -- without the simulations we have
carried out in this study. Therefore carrying out simulations on
a star-by-star basis is the \emph{only} way to understand the
selection effects in Doppler velocity planet searches.

\begin{figure}
  \begin{center}
    \leavevmode
    \epsfig{file=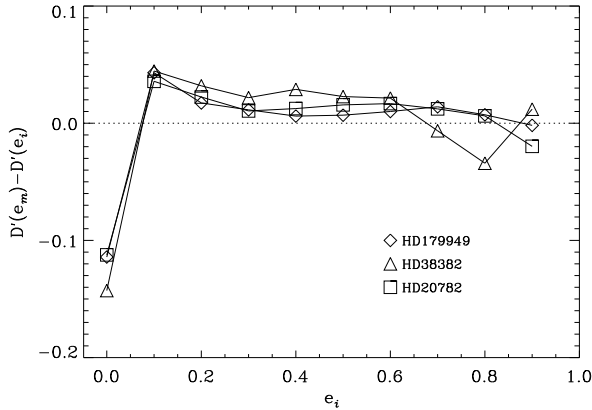,scale=0.47}
    \caption{The difference between measured and input detectabilities
      as function of input eccentricity. There are fewer measured
      detections at $e_i=0.0$ and a small excess at other
      eccentricities, peaking at $e_i=0.1$.}
    \label{fig:ediff}
  \end{center}
\end{figure}

\subsection{A Bias against Zero Eccentricity Detections}
\label{sub:ivm}

In the previous section, we examined $D^\prime_{\mathrm{int}}(e_i)$,
the detectability at each $e_i$, which is at all $P_i$ and
$M_i$. This is fine for the case of these
simulations, where we know the input parameter values \emph{a
  priori}. However, as this is never the case for actual Doppler planet
data it is useful to consider our detectability at each $e_m$; i.e. the
\emph{measured} eccentricities rather than the input eccentricities. We
call this quantity $D^\prime_{\mathrm{int}}(e_m)$. It is determined by
counting the number of correct detections -- i.e. false positives are
excluded -- in equally spaced bins of $e_m$, normalised
by the number of simulations in each of bin.

We now compare the two quantities, $D^\prime_{\mathrm{int}}(e_m)$ and
$D^\prime_{\mathrm{int}}(e_i)$ and show the difference between them
for each of our three stars in Figure \ref{fig:ediff}. The striking
feature is that when binned by measured eccentricity the detectability
is \emph{lower} by up to 15\% at $e_i=0$ than when binned by input
eccentricity. At other eccentricities $e_i$, there is an opposite
effect, albeit smaller. What could be causing these apparent biases,
especially against finding zero eccentricity orbits?

\begin{figure}
  \begin{center}
    \leavevmode
    \epsfig{file=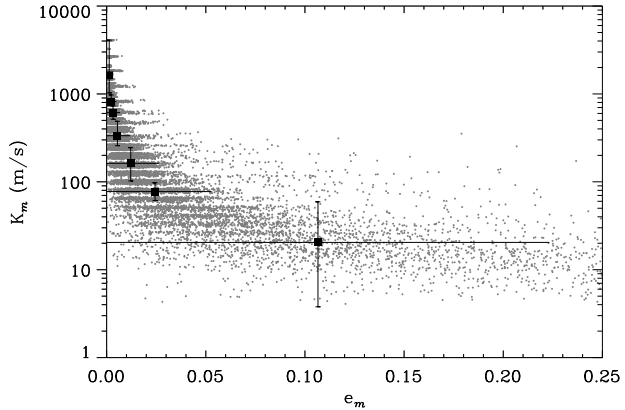,scale=0.47}
    \caption{Measured semi-amplitude $K_m$ as a function of measured
      eccentricity $e_m$ where $e_i=0.0$. Also plotted is the
      median value of $e_m$ over various ranges of $K_m$ (shown by the
      error bars in $K_m$) and the median value of the fit error for
      $e_m$ over the same range of $K_m$ (shown by the error bars in
      $e_m$). At low signal-to-noise ratios -- and therefore low
      values of $K_m$ -- there is a bias against measuring zero
      eccentricity orbits.}
    \label{fig:evK}
  \end{center}
\end{figure}

The answer can be seen in Figure \ref{fig:evK}, which shows the
measured semi-amplitude, $K_m$, as a function of the measured
eccentricity, $e_m$, for HD\,38382 where the input eccentricity is
$e_i=0.0$. Also plotted is the median value of $e_m$ over various
ranges of $K_m$ (shown by the error bars in $K_m$) and the median
value of the fit error for $e_m$ over the same range of $K_m$ (shown
by the error bars in $e_m$). As semi-amplitude decreases the measured
eccentricity ($e_m$) \emph{increases}, i.e.\ the Keplerian fit to
noisy data with a perfectly circular orbit ($e=0.0$) tends towards a
higher eccentricity.  This has the effect of decreasing the number of
$e_m=0$ orbits and increasing the number of non-zero eccentricity
orbits, in particular for the $e_m=0.1$ bin of orbits (Figure
\ref{fig:evK}). At low values of $K_m$, up to one third of zero
eccentricity orbits have best fits that move them out of the $e=0$
bin. This occurs because the \emph{shape} of the velocity curve is not well
constrained in low signal-to-noise data, even if the orbital period
and semi-amplitude are. For $e_i\ge 0.1$ eccentricity orbits the
distribution of measured eccentricities is symmetric so the median of
spread at low signal-to-noise converges to the real value and the
effect is not observed.  The distribution of measured eccentricities
is non-symmetric when $e_i \to 0.0$, however, so the median is always
non-zero.  Similar plots for the other two stars show the same effect
to varying degrees.

The conclusion from this analysis is that there is a small, but
significant, bias \emph{against} measuring an eccentricity of zero,
especially at low signal-to-noise ratios. This bias has also been
recently reported in an independent analysis by \citet{ST08}.

\subsection{Period}
\label{sub:period}

\begin{figure}
  \begin{center}
    \leavevmode
    \epsfig{file=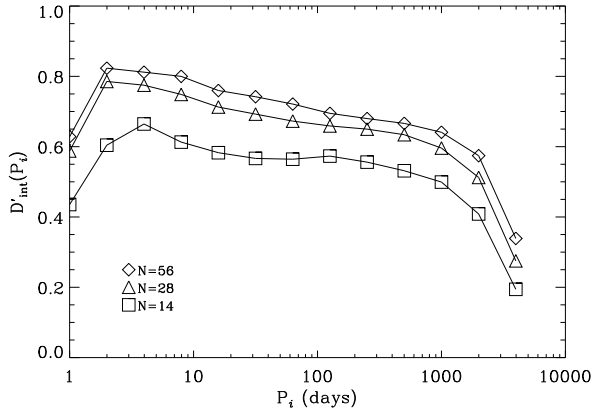,scale=0.47}
    \caption{Similar to Figure \ref{fig:eselfunc2} except as a
      function of input period.}
    \label{fig:Pselfunc_sub}
  \end{center}
\end{figure}

We now turn our attention to $D^\prime_{\mathrm{int}}(P_i)$, which is
the integrated detectability at a given input period, $P_i$. Figure
\ref{fig:Pselfunc_sub} shows $D^\prime_{\mathrm{int}}(P_i)$ for each
subset of the HD\,179949 data as a function of $P_i$. Perhaps
unsurprisingly, a higher density of observations leads to a higher
detectability of planets at all periods. For the 56 and 28 epoch data
sets, $D^\prime_{\mathrm{int}}(P_i)$ is a mostly linear function of $\log
P_i$, with a drop at $P_i \sim 2$\,d and a large drop-off $P_i\la
2$\,d (as noted above). For the 14 epoch subset, however,
$D^\prime_{\mathrm{int}}(P_i)$ drops sharply at $\la$4\,d, rather than
at $P_i\la 2$\,d, reflecting the fact that the reduced data set does
not sample short periods well.  At periods longer than $\Delta T$
(=\,2626 days for HD\,179949), $D^\prime_{\mathrm{int}}(P_i)$ drops
off by almost a factor of two.

\begin{figure}
  \begin{center}
    \leavevmode
    \psfig{file=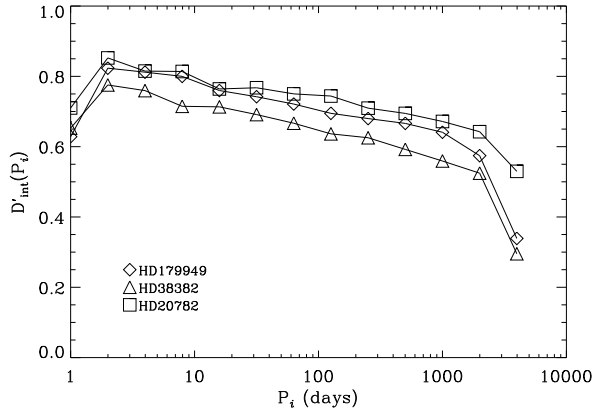,scale=0.47}
    \caption{Similar to Figure \ref{fig:eselfunc1} except as a
      function of input period.}
    \label{fig:Pselfunc_all}
  \end{center}
\end{figure}

We have seen the effects of sampling on the period detectability
above, but what role does the data quality (indicated by the median
measurement uncertainty in Table \ref{tab:targets}) play, if any? To
investigate this, we show $D^\prime_{\mathrm{int}}(P_i)$ for each of
the three objects studied in Figure \ref{fig:Pselfunc_all}. Once
again, we see a drop in detectability at periods below $\sim$\,2\,days
and at periods longer than $\Delta T$. While there is an offset
between the $D^\prime_{\mathrm{int}}(P_i)$ for each of the HD\,179949
subsets, caused by differing observation density, there is no clear
offset between the three different stars, despite the significantly
different observation densities (HD\,179949 -- 47\,days/epoch;
HD\,20782 -- 89\,days/epoch; HD\,38382 -- 144\,days/epoch). If examine
the median measurement uncertainties as a proxy for data quality, we
see for example that HD\,20782, which has the lowest measurement
uncertainties at 2.27\,m\,s$^{-1}$, has typically higher period
detectabilities, despite having lower observation density than HD\,179949.
This suggests that once again a complicated combination of observation
density and data quality are important in selection functions for
Doppler planet search data.

We also consider the detectability at each $P_m$, the measured
period, denoted $D^\prime_{\mathrm{int}}(P_m)$, and compare it with the
$D^\prime_{\mathrm{int}}(P_i)$ discussed above. We calculate
$D^\prime_{\mathrm{int}}(P_m)$ by counting the number of correct
detections in equally spaced bins of $\log P_m$, normalised
by the number of simulations in each of bin. At periods up to 1000\,d
($\log P=3.0$), the difference in detectabilities are less than
0.5\% for all stars. There is a small offset for the two longest
periods of up to $\pm$5\%, which is caused by poorly constrained long
periods ($\log P=3.6$) ``leaking'' into the next shorter period bin ($\log
P=3.3$).

\subsection{Planet mass}
\label{sub:M}

\begin{figure}
  \begin{center}
    \leavevmode
    \epsfig{file=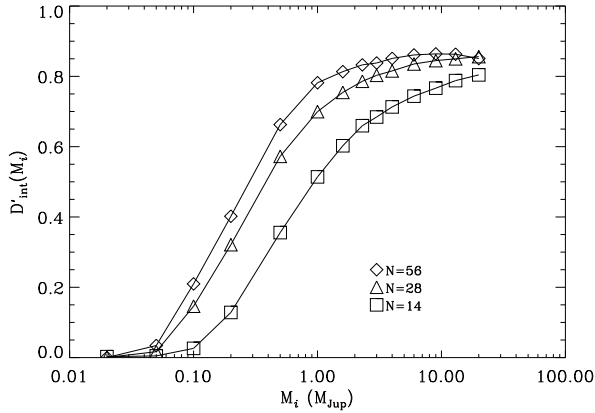,scale=0.47}
    \caption{Similar to Figure \ref{fig:eselfunc2}
      except as a function of input planet mass.}
    \label{fig:Mselfunc1}
  \end{center}
\end{figure}

The integrated detectability as a function of planet mass,
$D^\prime_{\mathrm{int}}(M_i)$ is more complicated than the integrated
detectability as a function of period or eccentricity, because planet
mass is a function of both of these parameters (as well as
semi-amplitude) through Equation \ref{eq:semiamp}. Figure
\ref{fig:Mselfunc1} shows $D^\prime_{\mathrm{int}}(M_i)$ as a function
of planet mass for the three HD\,179949 subsets. As one might naively
expect, we see that more data results in higher detectabilities for a
given mass of planet. (Recall also that at low
masses false positives begin to have a significant impact on false
positives for sparesly sampled data -- they
represent up to 20\% as a fraction of the total detections at
M$<$0.2\mjup, leading to an apparently higher detectability than data
sets with more observations.)

\begin{figure}
  \begin{center}
    \leavevmode
    \epsfig{file=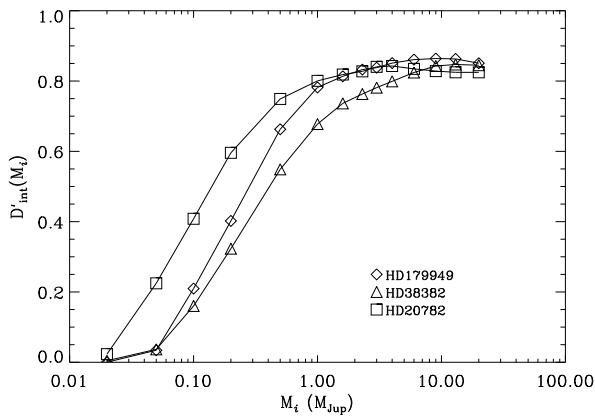,scale=0.47}
    \caption{Similar to Figure \ref{fig:eselfunc1}
      except for each of the three stars studied.}
    \label{fig:Mselfunc2}
  \end{center}
\end{figure}

Once again, if we examine the $D^\prime_{\mathrm{int}}(M_i)$ curves
for simulations of each star we find variations (see Figure
\ref{fig:Mselfunc2}). The HD\,20782 observations, which have the
highest quality with a median measurement uncertainty of
2.27\,m\,s$^{-1}$, allow the detection of the lowest planet masses
(after false positives are removed), as shown in Figure
\ref{fig:Mselfunc2}. Even though there are more
observations of HD\,179949, the median uncertainty of HD\,20782 is
less than half that star's value. In the case of HD\,38382, the median
uncertainty of the observations \emph{and} the number of epochs appear
to be important, giving a slightly lower number of planets detected at
intermediate masses  (1\,\mjup $< M_i < 6$\,\mjup) compared with the
HD\,179949. It is this complexity that shows the importance of a
star-by-star analysis.

Finally, we examine the detectability at $M_m$, the measured planet
mass, which we denote $D^\prime_{\mathrm{int}}(M_m)$, and once again
compare it with the corresponding detectability for input planet
mass. The detectabilities are binned in $\log M_m$, with the centroid
of each bin set to the corresponding value of $\log M_i$ to allow
comparison. The width of the bins was set to half the difference
between the adjacent bins. Recall that $D^\prime_{\mathrm{int}}(M_i)$
is calculated by counting the number of detections at a given $M_i$,
which assumes that the mass is known \emph{a priori}. We find the
difference between $D^\prime_{\mathrm{int}}(M_m)$ and
$D^\prime_{\mathrm{int}}(M_i)$ is within $\sim$3\% for each
star. We do not consider these differences to be significant.

\section{Summary}
\label{sec:conc}

We have begun a project to investigate the observational biases
inherent in Doppler velocity data, in particular in the
Anglo-Australian Planet Search. An essential part of this study is the
development of a set of tools that will allow the automatic detection
of exoplanets. We present the 2D Keplerian Lomb-Scargle periodogram, a
new algorithm based on an extension of the traditional Lomb-Scargle
periodogram, which includes eccentricity. This
is very efficient at detecting high eccentricity planets, which we
highlight with a re-analysis of the extreme object HD\,20782b. 

We have constructed Monte-Carlo-like simulations of Anglo-Australian
Planet Search data that include the time-stamps of the observations and
a simple noise model that incorporates into it the measurement
uncertainties. The simulations cover the full range of physically
important exoplanet
orbital parameters: period, eccentricity and planet mass.
As part of the simulation system, we have developed a set of detection
criteria which can be applied to our simulated data sets in an automated
fashion. We have investigated the false positives produced by
these detection criteria and
found them to be quantifiable and at an acceptably low level,
which enables meaningful conclusions to be reached from our
simulations.  We also find that there is a bias against
detecting zero-eccentricity orbits at low signal-to-noise ratios.

Finally, we present preliminary results from simulations of velocity
observations of three AAPS objects: HD\,179949, HD\,38382 and
HD\,20782. Our investigation shows that the detectability of exoplanet
orbital parameters differs significantly from object to object,
meaning that the simple parametrisations invoked by previous studies
are of limited validity.

\vspace{0.1cm}
We thank Rob Wittenmyer for commenting on the manuscript.
We gratefully acknowledge the superb technical support at the
Anglo-Australian Telescope which has been critical to the success of
this project.   We acknowledge support by PPARC
grant PPC/C000552/1 (SJOT), NSF grant AST-9988087 and travel support
from the Carnegie Institution of Washington (RPB), NASA grant
NAG5-8299 and NSF grant AST95-20443 (GWM), and ARC grant DP0774000 (CGT).   
We thank the Australian and UK Telescope assignment committees for generous
allocations of telescope time.  This research has made use of NASA's
Astrophysics Data System, and the SIMBAD database, operated at CDS,
Strasbourg, France. We would like to extend our effusive thanks to
Professor Matthew Bailes and the
staff at the Supercomputing Centre at the Swinburne University of
Technology for allowing us the use of their facilities, and providing
support and assistance when required. The authors acknowledge the use
of UCL Research Computing facilities and services in the completion of
this work.

\bibliographystyle{mn2e}
\bibliography{%
mnemonic,%
mnemonic-simple,%
planets,%
planets_stat,%
astero,%
stellar,%
tech%
}

\begin{thebibliography}{}

\bibitem[\protect\citeauthoryear{{Bevington}}{{Bevington}}{1969}]{Beving69}
{Bevington} P.~R.,  1969, Data Reduction and Error Analysis for the Physical
  Sciences.
McGraw-Hill, p.~72

\bibitem[\protect\citeauthoryear{{Butler}, {Marcy}, {Williams}, {McCarthy},
  {Dosanjh} \& {Vogt}}{{Butler} et~al.}{1996}]{BMW96}
{Butler} R.~P.,  {Marcy} G.~W.,  {Williams} E.,  {McCarthy} C.,  {Dosanjh} P.,
    {Vogt} S.~S.,  1996, PASP, 108, 500

\bibitem[\protect\citeauthoryear{{Butler}, {Tinney}, {Marcy}, {Jones}, {Penny}
  \& {Apps}}{{Butler} et~al.}{2001}]{BTM01}
{Butler} R.~P.,  {Tinney} C.~G.,  {Marcy} G.~W.,  {Jones} H.~R.~A.,  {Penny}
  A.~J.,    {Apps} K.,  2001, ApJ, 555, 410

\bibitem[\protect\citeauthoryear{{Butler}, {Wright}, {Marcy}, {Fischer},
  {Vogt}, {Tinney}, {Jones}, {Carter}, {Johnson}, {McCarthy} \&
  {Penny}}{{Butler} et~al.}{2006}]{BWM06}
{Butler} R.~P.,  {Wright} J.~T.,  {Marcy} G.~W.,  {Fischer} D.~A.,  {Vogt}
  S.~S.,  {Tinney} C.~G.,  {Jones} H.~R.~A.,  {Carter} B.~D.,  {Johnson} J.~A.,
   {McCarthy} C.,    {Penny} A.~J.,  2006, ApJ, 646, 505

\bibitem[\protect\citeauthoryear{{Cumming}}{{Cumming}}{2004}]{Cumming04}
{Cumming} A.,  2004, MNRAS, 354, 1165

\bibitem[\protect\citeauthoryear{{Cumming}, {Butler}, {Marcy}, {Vogt}, {Wright}
  \& {Fischer}}{{Cumming} et~al.}{2008}]{CBM08}
{Cumming} A.,  {Butler} R.~P.,  {Marcy} G.~W.,  {Vogt} S.~S.,  {Wright} J.~T.,
    {Fischer} D.~A.,  2008, PASP

\bibitem[\protect\citeauthoryear{{Cumming}, {Marcy} \& {Butler}}{{Cumming}
  et~al.}{1999}]{CMB99}
{Cumming} A.,  {Marcy} G.~W.,    {Butler} R.~P.,  1999, ApJ, 526, 890

\bibitem[\protect\citeauthoryear{{Ford}}{{Ford}}{2005}]{Ford05}
{Ford} E.~B.,  2005, AJ, 129, 1706

\bibitem[\protect\citeauthoryear{{Gaudi}, {Seager} \& {Mallen-Ornelas}}{{Gaudi}
  et~al.}{2005}]{GSM-O05}
{Gaudi} B.~S.,  {Seager} S.,    {Mallen-Ornelas} G.,  2005, ApJ, 623, 472

\bibitem[\protect\citeauthoryear{{H{\"o}gbom}}{{H{\"o}gbom}}{1974}]{Hoegbom74}
{H{\"o}gbom} J.~A.,  1974, A\&A, 15, 417

\bibitem[\protect\citeauthoryear{{Jones}, {Butler}, {Tinney}, {Marcy},
  {Carter}, {Penny}, {McCarthy} \& {Bailey}}{{Jones} et~al.}{2006}]{JBT06}
{Jones} H.~R.~A.,  {Butler} R.~P.,  {Tinney} C.~G.,  {Marcy} G.~W.,  {Carter}
  B.~D.,  {Penny} A.~J.,  {McCarthy} C.,    {Bailey} J.,  2006, MNRAS, 369, 249

\bibitem[\protect\citeauthoryear{{Jones}, {Paul Butler}, {Marcy}, {Tinney},
  {Penny}, {McCarthy} \& {Carter}}{{Jones} et~al.}{2002}]{JBM02}
{Jones} H.~R.~A.,  {Paul Butler} R.,  {Marcy} G.~W.,  {Tinney} C.~G.,  {Penny}
  A.~J.,  {McCarthy} C.,    {Carter} B.~D.,  2002, MNRAS, 337, 1170

\bibitem[\protect\citeauthoryear{{Lindman}}{{Lindman}}{1974}]{Lindman74}
{Lindman} H.~R.,  1974, {Analysis of Variance in Complex Experimental Designs}.
W.~H.~Freeman \& Co.

\bibitem[\protect\citeauthoryear{{Lomb}}{{Lomb}}{1976}]{Lomb76}
{Lomb} N.~R.,  1976, Ap\&SS, 39, 447

\bibitem[\protect\citeauthoryear{{Marcy}, {Butler}, {Vogt}, {Fischer}, {Henry},
  {Laughlin}, {Wright} \& {Johnson}}{{Marcy} et~al.}{2005}]{MBV05}
{Marcy} G.~W.,  {Butler} R.~P.,  {Vogt} S.~S.,  {Fischer} D.~A.,  {Henry}
  G.~W.,  {Laughlin} G.,  {Wright} J.~T.,    {Johnson} J.~A.,  2005, ApJ, 619,
  570

\bibitem[\protect\citeauthoryear{{Narayan}, {Cumming} \& {Lin}}{{Narayan}
  et~al.}{2005}]{NCL05}
{Narayan} R.,  {Cumming} A.,    {Lin} D.~N.~C.,  2005, ApJ, 620, 1002

\bibitem[\protect\citeauthoryear{{O'Toole}, {Butler}, {Tinney}, {Marcy},
  {Carter}, {McCarthy}, {Bailey} \& {Penny} A.~J.~{Apps}}{{O'Toole}
  et~al.}{2007}]{OBT07}
{O'Toole} S.~J.,  {Butler} R.~P.,  {Tinney} C.~G.,  {Marcy} G.~W.,  {Carter}
  B.,  {McCarthy} C.,  {Bailey} J.,    {Penny} A.~J.~{Apps} K.,  2007, ApJ,
  660, 1636

\bibitem[\protect\citeauthoryear{{O'Toole}, {Tinney} \& {Jones}}{{O'Toole}
  et~al.}{2008}]{OTJ08}
{O'Toole} S.~J.,  {Tinney} C.~G.,    {Jones} H. R.~A.,  2008, MNRAS, 386, 516

\bibitem[\protect\citeauthoryear{{Press}, {Flannery} \& {Teukolsky}}{{Press}
  et~al.}{1986}]{Press86}
{Press} W.~H.,  {Flannery} B.~P.,    {Teukolsky} S.~A.,  1986, {Numerical
  Recipes}.
Cambridge Univ. Press

\bibitem[\protect\citeauthoryear{{Scargle}}{{Scargle}}{1982}]{Scargle82}
{Scargle} J.~D.,  1982, ApJ, 263, 835

\bibitem[\protect\citeauthoryear{{Shen} \& {Turner}}{{Shen} \&
  {Turner}}{2008}]{ST08}
{Shen} Y.,  {Turner} E.~L.,  2008, ApJ

\bibitem[\protect\citeauthoryear{{Tinney}, {Butler}, {Marcy}, {Jones}, {Penny},
  {McCarthy}, {Carter} \& {Bond}}{{Tinney} et~al.}{2003}]{TBM03}
{Tinney} C.~G.,  {Butler} R.~P.,  {Marcy} G.~W.,  {Jones} H.~R.~A.,  {Penny}
  A.~J.,  {McCarthy} C.,  {Carter} B.~D.,    {Bond} J.,  2003, ApJ, 587, 423

\bibitem[\protect\citeauthoryear{{Tinney}, {Butler}, {Marcy}, {Jones}, {Penny},
  {McCarthy}, {Carter} \& {Fischer}}{{Tinney} et~al.}{2005}]{TBM05}
{Tinney} C.~G.,  {Butler} R.~P.,  {Marcy} G.~W.,  {Jones} H.~R.~A.,  {Penny}
  A.~J.,  {McCarthy} C.,  {Carter} B.~D.,    {Fischer} D.~A.,  2005, ApJ, 623,
  1171

\bibitem[\protect\citeauthoryear{{Tinney}, {Butler}, {Marcy}, {Jones}, {Penny},
  {Vogt}, {Apps} \& {Henry}}{{Tinney} et~al.}{2001}]{TBM01}
{Tinney} C.~G.,  {Butler} R.~P.,  {Marcy} G.~W.,  {Jones} H.~R.~A.,  {Penny}
  A.~J.,  {Vogt} S.~S.,  {Apps} K.,    {Henry} G.~W.,  2001, ApJ, 551, 507

\bibitem[\protect\citeauthoryear{{Valenti} \& {Fischer}}{{Valenti} \&
  {Fischer}}{2005}]{VF05}
{Valenti} J.~A.,  {Fischer} D.~A.,  2005, ApJS, 159, 141

\bibitem[\protect\citeauthoryear{{Wittenmyer}, {Endl}, {Cochran}, {Hatzes},
  {Walker}, {Yang} \& {Paulson}}{{Wittenmyer} et~al.}{2006}]{WEC06}
{Wittenmyer} R.~A.,  {Endl} M.,  {Cochran} W.~D.,  {Hatzes} A.~P.,  {Walker}
  G.~A.~H.,  {Yang} S.~L.~S.,    {Paulson} D.~B.,  2006, AJ, 132, 177

\bibitem[\protect\citeauthoryear{{Wright}}{{Wright}}{2005}]{Wright05}
{Wright} J.~T.,  2005, PASP, 117, 657

\end{thebibliography}

\end{document}